%% file: bm-casc-2021.tex
\documentclass{article}

\usepackage{cite}

\usepackage{geometry}
\input{settings}

\input{macros}

\usepackage{booktabs}

\newcommand{\email}[1]{\texttt{#1}}

\newif\iflongversion
\longversionfalse

\begin{document}

\renewcommand{\thelstlisting}{\arabic{lstlisting}}

\title{\textbf{\Large{On the Complexity and Parallel Implementation of Hensel's Lemma and Weierstrass Preparation}}}
\author{Alexander Brandt,  Marc Moreno~Maza \\
Department of Computer Science, The University of Western Ontario, London, Canada\\
\email{abrandt5@uwo.ca}, \email{moreno@csd.uwo.ca}
}
\date{}

\maketitle

\begin{abstract}
\input{abstract}
\\[0.5em]
\textbf{Keywords:} Formal power series $\cdot$ Weierstrass preparation $\cdot$ Hensel's lemma $\cdot$ Hensel factorization $\cdot$ Parallel processing $\cdot$ Parallel pipeline
\end{abstract}

\input{intro}

\input{background}
\input{lazy-powerseries}

\input{complexity}
\input{parallel}

\input{experimentation}

\subsubsection*{Acknowledgements}
The authors would like to thank NSERC of Canada (award CGSD3-535362-2019).

\renewcommand{\refname}{References}
\bibliographystyle{plain}
\bibliography{references}

\end{document}

%% file: settings.tex
\usepackage{authblk}
\usepackage{lhelp}
\usepackage{amsfonts,latexsym}
\usepackage{amsmath}
\usepackage{amssymb}
\usepackage{algorithm}
\usepackage{algorithmicx}
\usepackage{algpseudocode}
\usepackage{verbatim}
\def\algorithmicfontsize{\scriptsize}
\algrenewcommand\alglinenumber[1]{\scriptsize #1:}

\usepackage{multirow}

\usepackage{enumerate}
\usepackage{nicefrac}

\usepackage{listings}
\usepackage[hidelinks]{hyperref}
\usepackage{cleveref}
\usepackage{color}

\usepackage{graphicx}
\usepackage{pgfplots}
\usepackage{tikz}
\usetikzlibrary{positioning,calc}
\usetikzlibrary{arrows.meta,chains,decorations.pathreplacing,scopes,positioning}
\usetikzlibrary{calc,shapes.multipart,chains,arrows}
\graphicspath{ {figures/} }
\DeclareGraphicsExtensions{.eps,.png}

\usepackage{nicefrac}

\usepackage{etoolbox}
\usepackage{tikz}
\usetikzlibrary{tikzmark}
\usetikzlibrary{calc}

\usepackage{scrextend} %for add margin
\usepackage{etoolbox}
\usepackage{xcolor}
\definecolor{commentgreen}{RGB}{2,112,10}
\definecolor{eminence}{RGB}{158,58,180}
\definecolor{weborange}{RGB}{230,145,0}
\definecolor{frenchplum}{rgb}{0.55, 0.0, 0.55}

\newtoggle{InString}{}% Keep track of if we are within a string
\togglefalse{InString}% Assume not initially in string

\usepackage{listings}
\newcommand\digitstyle{\color{eminence}}
\makeatletter
\newcommand{\ProcessDigit}[1]
{%
	\ifnum\lst@mode=\lst@Pmode\relax%
	{\digitstyle #1}%
	\else
	#1%
	\fi
}
\makeatother
\lstdefinestyle{mystyle} {
	language=C++,
	frame=tb,
	tabsize=4,
	showstringspaces=false,
	numbers=left,
	backgroundcolor=\color{black!3}, % set backgroundcolor
	belowcaptionskip=1\baselineskip,
	breaklines=true,
	commentstyle=\color{commentgreen},
	keywordstyle=\bfseries\color{eminence},
	numberstyle=\tiny\color{gray},
	stringstyle=\color{weborange},
	basicstyle=\scriptsize\ttfamily, % basic font setting
	emph={int,char,double,float,unsigned,void,bool,AltArrZ\_t, mpz\_t, degrees\_t, SMZP, RationalNumber, HomogPartGenerator\_u, PowerSeries\_t, Poly\_ptr, HomogGenParamType\_e, integer},
	emphstyle={\color{blue!60!black}},
	otherkeywords={>,<,.,;,-,!,=,~, proc, then, end },%,1,2,3,4,5,6,7,8,9,0},
	morekeywords = [2]{>,<,.,;,-,!,=,~},
	morecomment=[l][\color{commentgreen}]{\#},
	keywordstyle = [2]\color{black},
}

\errorcontextlines\maxdimen

\newcommand{\ALGtikzmarkcolor}{black}% customise this, if you want
\newcommand{\ALGtikzmarkextraindent}{4pt}% customise this, if you want
\newcommand{\ALGtikzmarkverticaloffsetstart}{-1.0ex}% customise this, if you want
\newcommand{\ALGtikzmarkverticaloffsetend}{-0.3ex}% customise this, if you want
\makeatletter
\newcounter{ALG@tikzmark@tempcnta}

\newcommand\ALG@tikzmark@start{%
	\global\let\ALG@tikzmark@last\ALG@tikzmark@starttext%
	\expandafter\edef\csname ALG@tikzmark@\theALG@nested\endcsname{\theALG@tikzmark@tempcnta}%
	\tikzmark{ALG@tikzmark@start@\csname ALG@tikzmark@\theALG@nested\endcsname}%
	\addtocounter{ALG@tikzmark@tempcnta}{1}%
}

\def\ALG@tikzmark@starttext{start}
\newcommand\ALG@tikzmark@end{%
	\ifx\ALG@tikzmark@last\ALG@tikzmark@starttext
	\else
	\tikzmark{ALG@tikzmark@end@\csname ALG@tikzmark@\theALG@nested\endcsname}%
	\tikz[overlay,remember picture] \draw[\ALGtikzmarkcolor] let \p{S}=($(pic cs:ALG@tikzmark@start@\csname ALG@tikzmark@\theALG@nested\endcsname)+(\ALGtikzmarkextraindent,\ALGtikzmarkverticaloffsetstart)$), \p{E}=($(pic cs:ALG@tikzmark@end@\csname ALG@tikzmark@\theALG@nested\endcsname)+(\ALGtikzmarkextraindent,\ALGtikzmarkverticaloffsetend)$) in (\x{S},\y{S})--(\x{S},\y{E});%
	\fi
	\gdef\ALG@tikzmark@last{end}%
}

\apptocmd{\ALG@beginblock}{\ALG@tikzmark@start}{}{\errmessage{failed to patch}}
\pretocmd{\ALG@endblock}{\ALG@tikzmark@end}{}{\errmessage{failed to patch}}
\makeatother

%% file: macros.tex
\newcommand{\maple}{{\sc Maple}}

\newcommand{\myproof}{\textsc{Proof.} }
\newcommand{\myfoorp}{\hfill$\square$\medskip}

\newcounter{theorem}[section]
\newtheorem{Theorem}[theorem]{Theorem}

\newtheorem{Lemma}[theorem]{Lemma}
\newtheorem{Corollary}[theorem]{Corollary}
\newtheorem{Conjecture}[theorem]{Conjecture}

\newtheorem{Observation}[theorem]{Observation}
\numberwithin{theorem}{section}

\def\K {\ensuremath{\mathbb{K}}}

\def\N {\ensuremath{\mathbb{N}}}

\def\x {\ensuremath{\mathbf{x}}}

\def\y {\ensuremath{\mathbf{y}}}

\renewcommand{\hat}[1]{\mbox{$\widehat{#1}$}}

\renewcommand{\deg}[1]{\mbox{{\rm deg}$(#1)$}}

\usepackage{color, colortbl}
\definecolor{LRed}{rgb}{1,.8,.8}
\definecolor{MRed}{rgb}{1,.6,.5}
\definecolor{HRed}{rgb}{0.7,.5,.6}
\definecolor{HGray}{rgb}{0.8,.8,.8}
\definecolor{HGray2}{rgb}{0.9,.9,.9}
\definecolor{LBlue}{rgb}{0.8,1,1}

\newcolumntype{C}{>{\columncolor{LRed}}c}{\newcolumntype{C}{c}}
\newcolumntype{D}{>{\columncolor{MRed}}c}{\newcolumntype{D}{c}}
\newcolumntype{E}{>{\columncolor{HRed}}c}{\newcolumntype{E}{c}}
\newcolumntype{F}{>{\columncolor{HRed}}c}{\newcolumntype{F}{c}}
\newcolumntype{G}{>{\columncolor{HGray}}c}{\newcolumntype{G}{c}}
\usepackage{array}
\newcolumntype{?}{!{\vrule width 1pt}}

\makeatletter
\providecommand*{\cupdot}{%
	\mathbin{%
		\mathpalette\@cupdot{}%
	}%
}
\newcommand*{\@cupdot}[2]{%
	\ooalign{%
		$\m@th#1\cup$\cr
		\hidewidth$\m@th#1\cdot$\hidewidth
	}%
}
\makeatother

\newcommand{\hidetext}[1]{\mbox{ \ }}

\definecolor{nicedarkgreen}{rgb}{0.05, 0.35, 0.12}

\newcommand{\textblockcomment}[1]{}
\renewcommand{\gets}{:=}

\newcolumntype{?}{!{\vrule width 1pt}}

\newcommand{\parfor}{{\textbf{parallel\_for}}}
\algblockdefx[parafor]{ParaFor}{EndParaFor}[1]{\textbf{parallel\_for} #1}{\textbf{end for}}
\algblockdefx[workpile]{ParaWhile}{EndParaWhile}[1]{\textbf{in parallel while} #1}{\textbf{end while}}
\algdef{SE}[DOWHILE]{Do}{doWhile}{\algorithmicdo}[1]{\algorithmicwhile\ #1}%

\newcommand{\yield}{\textbf{yield}}

%% file: abstract.tex
Hensel's lemma, combined with repeated applications of Weierstrass preparation theorem, 
allows for the factorization of polynomials 
with multivariate power series coefficients.
We present a complexity analysis for this
method and leverage those results
to guide the load-balancing of a parallel implementation 
to concurrently update all factors. 
%We further discuss parallel processing techniques for
%simultaneously updating all of the factors.  
In particular, the factorization creates a \textit{pipeline} 
where the terms of degree $k$ of the first factor are computed simultaneously
with the terms of degree $k-1$ of the second factor, etc. 
An implementation challenge is the inherent irregularity 
of computational work between factors, 
as our complexity analysis reveals.
Additional resource utilization and load-balancing
is achieved through
the parallelization of Weierstrass preparation.
%This pipeline is then bolstered through the
%parallelization of Weierstrass preparation to improve 
%resource utilization and load-balancing.
Experimental results show the efficacy of this mixed parallel scheme, 
achieving up to 9$\times$ parallel speedup on a 12-core machine.

%% file: intro.tex
\section{Introduction}
\label{sec:intro}

Factorization via Hensel's lemma, or simply Hensel factorization,
provides a mechanism for factorizing univariate polynomials with multivariate power series coefficients.
%% In particular, for a monic square-free multivariate polynomial in $(X_1,\ldots,X_n,Y)$,
In particular, for a multivariate polynomial in $(X_1,\ldots,X_n,Y)$,
monic and square-free as a polynomial in $Y$, 
%or a univariate polynomial in $Y$ with multivariate power series coefficients in $X_1,\ldots,X_n$, 
one can compute its roots with respect to $Y$ as power series in $(X_1,\ldots,X_n)$.
For a bivariate polynomial in $(X_1,Y)$, the classical Newton--Puiseux method 
is known to compute the polynomial's roots with respect to $Y$ as univariate Puiseux series in $X_1$. 
The transition from power series to Puiseux series arises from handling 
the non-monic case.

The \textit{Hensel--Sasaki Construction} or \textit{Extended Hensel Construction} (EHC)
was proposed in \cite{Sasaki1999} as an efficient alternative to the
Newton--Puiseux method for the case of univariate coefficients. In the same paper, 
an extension of the Hensel--Sasaki construction for multivariate coefficients
was proposed, and then later extended, see e.g., \cite{iwami2003analytic,DBLP:conf/casc/SasakiI16}.
In \cite{DBLP:journals/jsc/AlvandiAKM20}, EHC
was improved in terms of algebraic complexity and practical implementation. 
%Moreover, the Hensel--Sasaski construction can be seen as a effective variation of 
%the Abhyunkar-Jung Theorem \cite[Theorem 1.3]{PARUSINSKI201229}.

In this paper, we present a parallel algorithm and its implementation for Hensel factorization
based on repeated applications of Weierstrass preparation theorem.
Our method uses a \textit{lazy evaluation} scheme,
meaning that more terms can be computed on demand without
having to restart the computation. This contrasts with a \textit{truncated} 
implementation where only terms up to a pre-deter\-mined degree are computed.
Unfortunately, such a degree often cannot be determined before calculations start,
or later may be found to not go far enough. This scenario occurs, for instance,
when computing limits of real rational functions~\cite{DBLP:journals/jsc/AlvandiAKM20}.

Lazy evaluation is not new, having previously been employed
in sparse polynomial arithmetic \cite{DBLP:conf/casc/MonaganV09}
and \textit{univariate} power series arithmetic
\cite{burge1987infinite,DBLP:journals/jsc/Hoeven02a}.
Our previous work in \cite{DBLP:conf/casc/BrandtKM20} is, to
the best of our knowledge, the first lazy multivariate power series implementation.
%%
%%%%
Our implementation of lazy and parallel power series
supports an arbitrary number of variables.
However, the complexity estimates of our proposed methods are measured
in the bivariate case; see Section~\ref{sec:complexity}.
This allows us to obtain sharp complexity estimates, 
giving the number of operations required to update each
factor of a Hensel factorization individually.
This information helps guide and load-balance our parallel implementation.
Further, limiting to the bivariate case allows for comparison 
with existing works.

Denote by $M(n)$ a
polynomial multiplication time~\cite[Ch. 8]{Gathen:2003:MCA:945759}
(the cost sufficient to multiply two polynomials of degree $n$),
Let $\K$ be algebraically closed and $f \in \K[[X_1]][Y]$ have degree $d_Y$ in $Y$ and total degree $d$.
Our Hensel factorization computes the first $k$ terms
of all factors of $f$ within $\mathcal{O}(d_Y^3k + d_Y^2k^2)$ operations in $\K$.
We conjecture in Section~\ref{sec:complexity}
that we can achieve $\mathcal{O}(d_Y^3k + d_Y^2M(k)\log k)$ using
relaxed algorithms \cite{DBLP:journals/jsc/Hoeven02a}.
%%%
The EHC of~\cite{DBLP:journals/jsc/AlvandiAKM20}
computes the first $k$ terms of all factors in $\mathcal{O}(d^3M(d) + k^2dM(d))$.
Kung and Traub show that, over the complex numbers $\mathbb{C}$,
the Newton--Puiseux method can do the same in
$\mathcal{O}(d^2kM(k))$ (resp. $\mathcal{O}(d^2M(k))$) operations in $\mathbb{C}$
using a linear lifting (resp. quadratic lifting) scheme \cite{DBLP:journals/jacm/KungT78}.
This complexity is lowered to $\mathcal{O}(d^2k)$
by Chudnovsky and Chudnovsky in~\cite{chudnovsky1986expansion}.
Berthomieu, Lecerf, and Quintin in \cite{DBLP:journals/aaecc/BerthomieuLQ13} also present an algorithm and implementation based on Hensel lifting 
which performs in $\mathcal{O}(M(d_Y)\log(d_Y)\,kM(k))$;
this is better 
than previous methods with respect to $d$ (or $d_Y$), but worse with respect to $k$. 

However, these estimates ignore an initial root finding step.
Denote by $R(n)$ the cost of finding the roots in $\K$ of a degree $n$ polynomial
(e.g. \cite[Th. 14.18]{Gathen:2003:MCA:945759}).
Our method then performs in 
$\mathcal{O}(d_Y^3k + d_Y^2k^2 + R(d_Y))$.
Note that the $R(d_Y)$ term does not depend on $k$, 
and is thus ignored henceforth.
For comparison, however,
Neiger, Rosenkilde, and Schost in \cite{DBLP:conf/issac/NeigerRS17}
present an algorithm based on Hensel lifting which,
\textit{ignoring polylogarithmic factors}, 
performs in $\mathcal{O}(d_Yk + kR(d_Y))$.%, but do not present an implementation.

Nonetheless, despite a higher asymptotic complexity, the formulation of EHC
in~\cite{DBLP:journals/jsc/AlvandiAKM20} is shown to be 
practically much more efficient than that of Kung and Traub.
Our serial implementation of lazy Hensel factorization
(using plain, quadratic arithmetic)
has already been shown in~\cite{DBLP:conf/casc/BrandtKM20}
to be orders of magnitude faster than that implementation of EHC.
Similarly, 
in \cite{DBLP:conf/casc/BrandtKM20}, we show that 
our serial lazy power series is orders of
magnitude faster than the truncated implementations 
of \textsc{Maple}'s \cite{maple} \texttt{mtaylor} and 
\textsc{SageMath}'s \cite{sagemath} \texttt{PowerSeriesRing}. 
This highlights that a lazy scheme using suboptimal
routines---but a careful implementation---can still be practically
efficient despite higher asymptotic complexity.

Further still, it is often the case that asymptotically fast 
algorithms are much more difficult to parallelize, 
and have high parallel overheads, e.g. polynomial multiplication based on FFT. 
Hence, in this work, we look to improve
the practical performance (i.e. when $k \gg d$)
of our previous lazy implementation 
through the use of parallel processing rather than by
reducing asymptotic bounds. 

In Hensel factorization, computing power series terms of each factor relies
on the computed terms of the previous factor. 
In particular, the output of one Weierstrass preparation becomes
the input to another. 
These successive dependencies naturally lead to a parallel \textit{pipeline}
or chain of \textit{producer-consumer} pairs.
Within numerical linear algebra,
pipelines have already been employed
in parallel implementations of singular value decomposition~\cite{DBLP:conf/sc/HaidarKL13},
LU decomposition, and Gaussian elimination~\cite{DBLP:journals/jcam/MichailidisM11}.
Meanwhile, to the best of our knowledge, the only use of parallel pipeline 
in symbolic computation is~\cite{DBLP:conf/issac/AsadiBMMX20}, which examines a
parallel implementation of triangular decomposition of polynomial systems.

However, in our case, work reduces with 
each pipeline stage, limiting throughput.
To overcome this challenge, we first make use of 
our complexity estimates to dynamically estimate the work required
to update each factor. Second, we compose parallel schemes
by applying the celebrated map-reduce pattern within
Weierstrass preparation, and thus within a stage of the pipeline.
Assigning multiple threads to a single pipeline stage improves 
load-balance and increases throughput.
Experimental results show this composition is effective, 
with a parallel speedup of up to $9\times$ on a 12-core machine.

The remainder of this paper is organized as follows.
Section~\ref{sec:bg} reviews mathematical background and notations.
Further background on our lazy power series of \cite{DBLP:conf/casc/BrandtKM20}
is presented in Section~\ref{sec:lazy-ps}.
Algorithms and complexity analyses of Weierstrass preparation
and Hensel factorization are given in Section~\ref{sec:complexity}.
Section~\ref{sec:parallel} presents our parallel variations, 
where our complexity estimates are used for dynamic scheduling.
Finally, Section~\ref{sec:exeripmentation} discusses experimental data.

%% file: background.tex
\section{Background}
\label{sec:bg}

We take this section to present basic concepts and notation of
multivariate power series and univariate polynomials over power series (UPoPS).
Further, we present constructive proofs for the theorems
of Weierstrass preparation and Hensel's lemma for UPoPS, 
from which algorithms are adapted; 
see Sections~\ref{sec:wpt-complexity} and \ref{sec:hensel-complexity}.
Further introductory details may be found in the book of
G. Fischer~\cite{GF01}. 
%Further implementation details may be found in \cite{DBLP:conf/casc/BrandtKM20}.
%%

\subsection{Power Series and Univariate Polynomials over Power Series}
\label{sec:powerseries-theory}

Let ${\K}$ be an algebraically closed
field. We denote by ${\K}[[X_1, \ldots, X_n]]$ the ring of
formal power series with coefficients
in {\K} and with variables $X_1, \ldots, X_n$.

Let $f = {\sum}_{e \in {{\N}^n}} \, a_e X^e$ be a formal power series,
where $a_e \in {\K}$, $X^e = X_1^{e_1}\cdots X_n^{e_n}$, $e = (e_1,\ldots,e_n) \in \mathbb{N}^n$,
and $|e| = e_1 + \cdots + e_n$.
Let $k$ be a non-negative integer.
\iflongversion
The {\it homogeneous part} and {\it polynomial part}
of $f$ in degree $k$ are denoted $f_{(k)}$ and $f^{(k)}$, and are 	
defined by
$f_{(k)} = {\sum}_{|e|=k} \, a_e X^e \quad
 \hbox{and} \quad
f^{(k)} = {\sum}_{i \leq k} \, f_{(i)}.$
\else
The {\it homogeneous part} 
of $f$ in degree $k$, denoted $f_{(k)}$, 
is defined by
$f_{(k)} = {\sum}_{|e|=k} \, a_e X^e$.
\fi
The {\it order} of $f$, denoted ${\rm ord}(f)$, 
is defined as ${\rm min} \{  i \mid f_{(i)} \neq 0 \}$, if $f \neq 0 $,
and as $\infty $ otherwise.

Recall several properties regarding power series. 
First, ${\K}[[X_1, \ldots, X_n]]$ is an integral domain.
Second, the set ${\cal M} = \{ f \in {\K}[[X_1, \ldots, X_n]] \mid {\rm ord}(f)  \geq 1 \}$ is the only maximal ideal of ${\K}[[X_1, \ldots, X_n]]$.
Third, for all $k \in {\N}$, we have ${\cal M}^k = \{ f \in {\K}[[X_1, \ldots, X_n]] \mid {\rm ord}(f)  \geq k \}$. Note that for 
$n=0$ we have $\mathcal{M} = \langle0\rangle$. 
Further, note that $f_{(k)} \in \mathcal{M}^k \setminus \mathcal{M}^{k+1}$
and $f_{(0)} \in \K$.
Fourth, a unit $u \in \K[[X_1,\ldots,X_n]]$ has ${\rm ord}(u) = 0$ or, equivalently, $u \not\in \mathcal{M}$.

Let $f, g, h, p \in \K[[X_1, \ldots, X_n]]$.
The {\it sum} and {\it difference} $f = g \pm h$
is given by ${\sum}_{k \in {\N}} \, (g_{(k)} \pm h_{(k)})$.
The product $p = g\,h$ is given by ${\sum}_{k \in {\N}}  \left( {\Sigma}_{i+ j = k} \, g_{(i)}  h_{(j)} \right).$
Notice that the these formulas naturally suggest a \textit{lazy evaluation}
scheme, where the result of an arithmetic operation
can be incrementally computed for increasing \textit{precision}.
% or, 
%equivalently, for homogeneous parts of increasing degree.
A power series $f$ is said to be known to precision $k \in \N$,
when $f_{(i)}$ is known for all $0 \leq i \leq k$.
Such an update function, parameterized by $k$,
for addition or subtraction 
is simply $f_{(k)} = g_{(k)} \pm h_{(k)}$;
an update function for multiplication is $p_{(k)} = \sum_{i=0}^k g_{(i)} h_{(k-i)}$.
Lazy evaluation is discussed further in Section~\ref{sec:lazy-ps}.
From these update formulas, the following observation follows.

\begin{Observation}[power series arithmetic]
\label{obs:ps-mult-ops}
Let $f, g, h, p \in \K[[X_1]]$ with $f = g \pm h$ and $p = g\,h$. $f_{(k)} = g_{(k)} \pm h_{(k)}$ can be computed in 1 operation in $\K$. $p_{(k)} = \sum_{i=0}^k g_{(i)} h_{(k-i)}$ can be computed in $2k -1$ operations in $\K$.
\end{Observation}
\iffalse
\myproof
For any power series in $\K[[X_1]]$, any of its homogeneous parts
belong to $\K$.
Computing $f_{(k)}$ is thus only one operation in $\K$.
Computing $p_{(k)}$ requires $k$ multiplications and $k-1$ additions
in $\K$.
\myfoorp
\fi

Now, let $f,g \in \mathbb{A}[Y]$ 
be univariate polynomials over power series where
$\mathbb{A} = \K[[X_1,\ldots,X_n]]$.
Writing $f = \sum_{i=0}^d a_iY^i$, for $a_i \in \mathbb{A}$ and $a_d \neq 0$, 
we have that the degree of $f$ (denoted $\deg{f,Y}$ or simply $\deg{f}$) is $d$.
%and its leading coefficient (denoted $\lc{f}$) is $a_d$.
Note that arithmetic operations for UPoPS are
%% easily inherited
easily derived 
from the arithmetic of its power series coefficients.
%Let $f$ and $g$ be of equal degree, appending zero terms to the one of lower degree otherwise.
%Further, let $g = \sum_{i=0}^d b_iY^i$.
%Then, $f + g = \sum_{i=0}^d (a_i + b_i)Y^i$ and $f\, g = \sum_{i=0}^{2d}\left(\sum_{j + \ell = i} a_j\,b_{\ell}\right)Y^i$.
%%
A UPoPS is said to be known up to precision $k$ if each of its power series
coefficients are known up to precision $k$.
A UPoPS $f$ is said to be \textit{general (in Y) of order $j$} if 
$f$ mod $\mathcal{M}[Y]$ has order $j$ when viewed as a power series in $Y$. 
%% That is, for $f = \sum_{i=0}^d a_iY^i$, $a_i \in \mathcal{M}$ for $0 \leq i < j$.
Thus, for $f \not\in \mathcal{M}[Y]$, writing
$f = \sum_{i=0}^d a_iY^i$, we have $a_i \in \mathcal{M}$ for $0 \leq i < j$
and $a_j \not\in  \mathcal{M}$.

\subsection{Weierstrass Preparation Theorem and Hensel Factorization}
%\newline Factorization via Hensel's Lemma}
\label{sec:weierstrass-bg}

The Weierstrass Preparation Theorem (WPT) 
is fundamentally a theorem regarding factorization. 
In the context of analytic functions, 
WPT implies that any analytic function 
resembles a polynomial in the neighbourhood of the origin.
Generally, WPT can be stated
for power series over power series, i.e. 
%for the power series $\K[[X_1,\ldots,X_n]][[Y]] = 
$\mathbb{A}[[Y]]$.
This can be used to prove that 
$\mathbb{A}$ is both a unique factorization domain
and a Noetherian ring. See \cite{DBLP:conf/casc/BrandtKM20} 
for such a proof of WPT.
Here, it is sufficient to state the theorem for UPoPS.
First, we begin with a simple lemma.
%which serves
%as the basis of our eventual proof of WPT and our implementation.
%%
\begin{Lemma}
\label{lemma:l4WPT}
Let $f,g,h \in \K[[X_1,\ldots,X_n]]$ such that $f= gh$. 
Let $f_i = f_{(i)}, g_i = g_{(i)}, h_i = h_{(i)}.$
If $f_0 = 0$ and $h_0 \neq 0$, then $g_k$ is uniquely
determined by $f_1,\ldots,f_k$ and $h_0,\ldots,h_{k-1}$
\end{Lemma}
\myproof
We proceed by induction on $k$. Since $f_0 = g_0h_0 = 0$ and $h_0 \neq 0$ both hold, the statement holds for $k=0$. Now let $k > 0$, 
assuming the hypothesis holds for $k-1$. 
To determine $g_k$ it is sufficient to expand $f = gh$ modulo $\mathcal{M}^{k+1}$: 
%\begin{align*}
$
f_1 + f_2 + \cdots + f_k = g_1h_0 + (g_1h_1 + g_2h_0) + \cdots + (g_1h_{k-1} + \cdots + g_{k-1}h_1 + g_kh_0);
$
%\end{align*}
and, recalling $h_0 \in \K \setminus \{0\}$, we have 
%\begin{align*}
%f_k = g_1h_{k-1} + \cdots + g_{k-1}h_1 + g_kh_0 \\
$
g_k = \nicefrac{1}{h_0} \left(f_k - g_1h_{k-1} - \cdots - g_{k-1}h_1\right)
$
.
%\end{align*}
\myfoorp

\begin{Theorem}[Weierstrass Preparation Theorem]
  \label{theorem:WPT}
  Let $f$ be a polynomial of $\K[[X_1,\ldots,X_n]][Y]$
  so that $f \not\equiv 0 \mod \mathcal{M}[Y]$ holds.
  Write $f = \sum_{i=0}^{d+m} a_iY^i$, with $a_i \in \K[[X_1,\ldots,X_n]]$,  
where $d \geq 0$ is the smallest integer such that
$a_d \not\in \mathcal{M}$ and $m$ is a non-negative integer.
Assume $f \not\equiv 0 \mod \mathcal{M}[Y]$.
Then, there exists a unique pair $p,\alpha$ satisfying the following: 
\begin{enumerate}[$(i)$]
	\item $f = p\, \alpha$,
	\item $\alpha$ is an invertible element of $\K[[X_1,\ldots,X_n]][[Y]]$,
	\item $p$ is a monic polynomial of degree $d$,
	\item writing $p = Y^d + b_{d-1}Y^{d-1} + \cdots b_1Y + b_0$, we have $b_{d-1},\ldots,b_0 \in \mathcal{M}$.
\end{enumerate}
\end{Theorem}
\myproof
If $n=0$, writing $f = \alpha Y^d$ with $\alpha = \sum_{i=0}^{m}a_{i+d}Y^i$ proves the existence of the decomposition.
Now, assume $n \geq 1$. Write $\alpha = \sum_{i=0}^m c_iY^i$, with $c_i \in \K[[X_1,\ldots,X_n]]$. 
We will determine $b_0,\ldots,b_{d-1},c_0,\ldots,c_m$ modulo successive powers of $\mathcal{M}$.
Since we require $\alpha$ to be a unit, $c_0 \not\in \mathcal{M}$
by definition. $a_0,\ldots,a_{d-1}$ are all 0 mod $\mathcal{M}$.  
Then, equating coefficients in $f = p\,\alpha$ we have: 
\begin{align}
\begin{array}{rcl}
a_0  & = & b_0 c_0 \\
a_1  & = &  b_0 c_1 + b_1 c_0 \\
%   a_2  & = &  b_0 c_2 + b_1 c_1 + b_2 c_0 \\
& \vdots &                         \\
a_{d-1} & = &  b_0 c_{d-1} + b_1 c_{d-2} + \cdots + b_{d-2} c_1 + b_{d-1} c_0 \\
a_d      & = &  b_0 c_{d} + b_1 c_{d-1} + \cdots + b_{d-1} c_1 + c_0 \\
%   a_{d+1}  & = & b_0 c_{d+1} + b_1 c_{d} + \cdots + b_{d-1} c_2 + c_1 \\
& \vdots &  \\  
%   a_{d+m-3} & = & b_{d-3}c_{m} + b_{d-2}c_{m-1} + b_{d-3}c_{m-2} + c_{m-3} \\
%   a_{d+m-2} & = & b_{d-2}c_{m} + b_{d-1}c_{m-1} + c_{m-2} \\
a_{d+m-1} & = & b_{d-1}c_{m} + c_{m-1} \\
a_{d+m} & = & c_m 
\end{array}\label{eqn:weier-eqs-upops}
\end{align}
and thus $b_0,\ldots,b_{d-1}$ are all 0 mod $\mathcal{M}$.
Then, $c_i \equiv a_{d+i} \mod \mathcal{M}$ for all $0 \leq i \leq m$.
All coefficients have thus been determined mod $\mathcal{M}$.
Let $k \in \mathbb{Z}^+$. Assume inductively that all 
$b_0,\ldots,b_{d-1},c_0,\ldots,c_m$ have been determined mod $\mathcal{M}^k$. 
%we will now determine them mod $\mathcal{M}^{k+1}$.

It follows from Lemma~\ref{lemma:l4WPT} that $b_0$ can be determined mod 
$\mathcal{M}^{k+1}$ from the equation $a_0 = b_0c_0$.
Consider now the second equation. Since $b_0$ is known
mod $\mathcal{M}^{k+1}$, and $b_0 \in \mathcal{M}$, the product $b_0c_1$ is also known mod $\mathcal{M}^{k+1}$.
%, despite $c_1$ only being known mod $\mathcal{M}^k$.
Then, we can determine $b_1$ using Lemma~\ref{lemma:l4WPT} and 
the formula $a_1 -  b_0c_1 = b_1c_0$. 
This procedure follows for $b_2,\ldots,b_{d-1}$.
With $b_0,\ldots,b_{d-1}$ known mod $\mathcal{M}^{k+1}$ 
%and, again, $b_0,\ldots,b_{d-1} \in \mathcal{M}$, 
each $c_0,\ldots,c_m$ can be determined mod $\mathcal{M}^{k+1}$ 
from the last $m+1$ equations. 
\myfoorp

One requirement of WPT is that
$f \not\equiv 0 \mod \mathcal{M}[Y]$. That is to say, 
$f$ cannot vanish at $(X_1,\ldots,X_n) = (0,\ldots,0)$ and, specifically, 
$f$ is general of order $d = \deg{p}$.
A suitable linear change in coordinates can always be applied
to meet this requirement; see Algorithm~\ref{alg:lincombgen} in 
Section~\ref{sec:complexity}.
Since Weierstrass preparation
provides a mechanism to factor a UPoPS into two factors, 
suitable changes in coordinates and several
applications of WPT can fully factorize a UPoPS.
The existence of such a factorization 
is given by Hensel's lemma for UPoPS.
%, and its proof 
%shows the construction of the factorization via WPT.
%This proof is transformed into an algorithm later in
%Section~\ref{sec:complexity}.
%
%For $f \in \K[[X_1,\ldots,X_n]][Y]$, define $\bar{f} = f(0,\ldots,0,Y) \in \K[Y]$.
%In this section we assume $\K$ is algebraically closed.

\begin{Theorem}[Hensel's Lemma]
\label{theorem:hensel}
Let $f = Y^d + \sum_{i=0}^{d-1} a_iY^i$ be a monic polynomial with $a_i \in \K[[X_1,\ldots,X_n]]$.
Let  $\bar{f} = f(0,\ldots,0,Y) = (Y-c_1)^{d_1}(Y-c_2)^{d_2}\cdots(Y-c_r)^{d_r}$ for $c_1,\ldots,c_r \in \K$
and positive integers $d_1,\ldots,d_r$. Then, there exists $f_1,\ldots,f_r \in \K[[X_1,\ldots,X_n]][Y]$, 
all monic in Y, such that:
\begin{enumerate}[$(i)$]
\item $f = f_1\cdots f_r$,
\item $\deg{f_i,Y} = d_i$ for $1 \leq i \leq r$, and
\item $\bar{f_i} = (Y-c_i)^{d_i}$ for $1 \leq i \leq r$.
\end{enumerate}
\end{Theorem} 
\myproof
We proceed by induction on $r$. For $r=1$, $d_1=d$ and we have $f_1 = f$,
where $f_1$ has all the required properties.
Now assume $r > 1$. A change of coordinates in $Y$, sends $c_r$ to 0. 
Define $g(X_1,\ldots,X_n,Y) = f(X_1,\ldots,X_n,Y+c_r) =
(Y+c_r)^d + a_{d-1}(Y+c_r)^{d-1} + \cdots + a_0$.
By construction, $g$ is general of order $d_r$
and WPT can be applied to obtain 
$g=p\,\alpha$ with $p$ being of degree $d_r$ and $\bar{p} = Y^{d_r}$.
Reversing the change of coordinates 
we set $f_r = p(Y-c_r)$ and $f^* = \alpha(Y-c_r)$,
and we have $f=f^*f_r$.
$f_r$ is a monic polynomial of degree $d_r$ in $Y$
with $\bar{f_r} = (Y-c_r)^{d_r}$ .
Moreover, we have $\bar{f^*} = (Y-c_1)^{d_1}(Y-c_2)^{d_2}\cdots(Y-c_{r-1})^{d_{r-1}}$.
The inductive hypothesis applied to $f^{*}$ implies the existence
of $f_1,\ldots,f_{r-1}$.
\myfoorp

%Notice that factorizing a UPoPS via Hensel's lemma and Weierstrass preparation 
%is equivalent to the \textit{Hensel-Sasaki Construction}, also known
%as the \textit{Extended Hensel Construction}
%\cite{Sasaki1999,DBLP:journals/jsc/AlvandiAKM20} when
%the polynomial (the UPoPS) to factor
%is monic in $Y$.

\subsection{Parallel Patterns}
\label{sec:parallel-patterns}

%Throughout this paper
We are concerned with \textit{thread-level parallelism},
where multiple threads of execution within a single process
enable concurrent processing.
Our parallel implementation 
%of Weierstrass preparation and Hensel factorization 
employs several so-called \textit{parallel patterns}---algorithmic structures
and organizations for efficient parallel processing.
We review a few patterns:
\textit{map}, \textit{producer-consumer}, and \textit{pipeline}.
See \cite{mccool2012structured}
for a detailed discussion.

%We are thus concerned with the issues of \textit{parallel overheads}
%such as the cost of spawning threads and \textit{over-subscription}---where
%the number of active software threads is greater than the number of hardware threads.
%These overheads can be reduced through the use of
%the aforementioned parallel patterns.

\iflongversion
\subsubsection{Map}
\else
\textbf{Map.}
\fi
The map pattern applies a function to each item in a collection, 
simultaneously executing the function on each independent data item.
Often, the application of a map produces a new collection
with the same shape as the input collection. Alternatively, 
the map pattern may modify each data item in place or, 
when combined with the \textit{reduce} pattern, produce 
a single data item. The reduce pattern 
combines data items pair-wise using some \textit{combiner} function.
%which takes two data items as input and outputs a single one.

When data items to be processed outnumber available threads, 
the map pattern can be applied block-wise, where
the data collection is (evenly) partitioned and each thread 
assigned a partition rather than a single data item.

Where a \textbf{for} loop has independent iterations, 
the map pattern is easily applied to execute each iteration of the loop
concurrently. Due to this ubiquity, the map pattern is often implicit
with such parallel for loops simply being labelled \parfor. 
In this way, the number of threads to use and
the partitioning of the data collection can be a dynamic
property of the algorithm.

\iflongversion
\subsubsection{Producer-Consumer and Asynchronous Generators}
\else
\textbf{Producer-Consumer and Asynchronous Generators.}
\fi
The producer-consumer pattern describes two functions
connected by a queue. The producer 
creates data items, pushing them to the queue, meanwhile 
the consumer processes data items, pulling them from the queue. 
Where both the creation of data and its processing
requires substantial work, producer
and consumer may operate concurrently, 
with the queue providing inter-thread communication.
%and stalling as necessary when the queue is empty or full.

A \textit{generator} or \textit{iterator}
is a special kind of co-routine function which \textbf{yield}s
data elements one at a time, rather than many together
as a collection; see, e.g. \cite[Ch. 8]{DBLP:books/daglib/0028503}.
Combining the producer-consumer pattern with generators 
allows for an \textit{asynchronous generator}, 
where the generator function is the producer
and the calling function is the consumer.
The intermediary queue allows the generator 
to produce items meanwhile the calling function
processes them.

\iflongversion
\subsubsection{Pipeline}
\else
\textbf{Pipeline.}
\fi
The pipeline pattern 
%(cf. pipelining from 
%instruction-level parallelism \cite[Ch. 3]{DBLP:books/daglib/0028244}) 
is a sequence of stages, where the output of one stage
is used as the input to another. Two consecutive stages
form a producer-consumer pair, with internal stages
being both a consumer and a producer.
Concurrency arises where each stage of the pipeline
may be executed in parallel.
Moreover, the pipeline pattern allows for earlier data items 
to flow from one stage to the next without waiting 
for later items to become available.
%Parallelism is best exploited where the time 
%to execute each stage is equal.

In terms of the latency of processing a single data item,
a pipeline does not improve upon its serial counterpart.
Rather, a parallel pipeline improves throughput, 
the amount of data that can be processed in a given amount of time.
Throughput is limited by the slowest stage of a pipeline, 
and thus special care must be given to ensure 
each stage of the pipeline runs in nearly equal time.

%A pipeline may be created statically, where
%each processing stage or function is programmed statically, 
%with the data items to process generated dynamically.
%However, 
A pipeline may be implicitly and dynamically 
created where an asynchronous generator 
consumes data from another asynchronous generator.
The number of asynchronous generator calls, and
thus the number of stages in the pipeline, can be 
dynamic to fit the needs of the application at runtime. 

%% file: lazy-powerseries.tex
\section{Lazy Power Series}
\label{sec:lazy-ps}

%Let $f,g,h \in \K[[X_1,\ldots,X_n]]$ be multivariate power series.
As we have seen in Section~\ref{sec:powerseries-theory}, 
certain arithmetic operations on power series naturally lead to a 
lazy evaluation scheme. 
In this scheme, homogeneous parts of a power series
are computed one at a time for increasing degree, as requested.
Our serial implementation of lazy power series
is detailed in \cite{DBLP:conf/casc/BrandtKM20}.
The underlying implementation
of (sparse multivariate) polynomial arithmetic 
is that of \cite{asadi2019algorithms}
%\footnote{
(indeed, dense multivariate arithmetic could prove beneficial, but that is left
to future work).
For the remainder of this paper, it is sufficient to
understand that lazy power series rely on 
the following three principles:
\begin{enumerate}[($i$)]
	\item an update function to compute the homogeneous part of a given degree;
	\item capturing of parameters required for that update function; and
	\item storage of previously computed homogeneous parts.
\end{enumerate}

Where a power series is constructed from arithmetic operations on other power series, 
the latter may be called the \textit{ancestors} of the former.
For example, the power series $f = g\,h$ has ancestors $g$ and $h$
and an update function $f_{(k)} = \sum_{i=0}^k g_{(i)}h_{(k-i)}$.
In implementation, and in the algorithms which follow in this paper, 
we can thus augment a power series with:
$(i)$ its current precision;
$(ii)$ references to its ancestors, if any; and 
$(iii)$ a reference to its update function.

Under this scheme, we make three remarks. 
Firstly, a power series can be lazily constructed
using essentially no work. Indeed, the initialization of a lazy power series
only requires specifying the appropriate update function 
and storing references to its ancestors. 
Secondly, specifying an update function and the ancestors of a power series
is sufficient for defining and computing that power series.
Thirdly, when updating a particular power series, 
its ancestors can automatically and recursively be updated as necessary
using their own update functions.

Hence, it is sufficient to simply define the update function
of a power series.
For example, Algorithm~\ref{alg:weierstrass-update} 
simultaneously updates $p$ and $\alpha$ as produced from a Weierstrass preparation.
Further, operations on power series should be understood to be only
the initialization of a power series, with no terms of the
power series yet computed; e.g., Algorithm~\ref{alg:hensel-factor} for Hensel factorization.

%% file: complexity.tex
\section{Algorithms and Complexity}
\label{sec:complexity}

In this section we present algorithms
for Weierstrass preparation and Hensel factorization
adapted from their constructive proofs; see Section~\ref{sec:bg}.
For each algorithm we analyze its complexity.
The algorithms---and eventual parallel variations and implementations, see Sections~\ref{sec:parallel}--\ref{sec:exeripmentation}---are
presented for the general multivariate case, with only the complexity
estimates limited to the bivariate case.
These results culminate
as Theorem~\ref{theorem:weierstrass-complexity}
and Corollary~\ref{coro:hensel-general-roots},
which respectively give the overall complexity of our algorithms for
WPT and Hensel factorization.
Meanwhile, Observation~\ref{obs:weierstrass-phase2}, Corollary~\ref{coro:wpt-monic}, and Theorem~\ref{theorem:hensel-per-factor}
more closely analyze the distribution of work
%within the algorithms
to guide and load-balance our parallel algorithms.
%described in Section~\ref{sec:parallel}.

\subsection{Weierstrass Preparation}
\label{sec:wpt-complexity}

From the proof of Weierstrass preparation (Theorem~\ref{theorem:WPT}),
we derive \textsc{WeierstrassUpdate} (Algorithm~\ref{alg:weierstrass-update}).
That proof proceeds modulo increasing powers of the maximal ideal $\mathcal{M}$,
which is equivalent to computing homogeneous parts of increasing degree,
just as required for our lazy power series.
For an application of Weierstrass preparation producing $p$ and $\alpha$,
this \textsc{WeierstrassUpdate} acts as the update
function for $p$ and $\alpha$, updating both simultaneously.

{
\begin{algorithm}[bth]
	\begin{algorithmic}[1]
		\algnotext{EndFor}
		\algnotext{EndIf}
		\algorithmicfontsize
		\caption{\small{\sc WeierstrassUpdate}($k$, $f$, $p$, $\alpha$)}\label{alg:weierstrass-update}
		\Require{$f = \sum_{i=0}^{d+m} a_iY^i$, $p = Y^d + \sum_{i=0}^{d-1} b_iY^i$, $\alpha = \sum_{i=0}^m c_iY^i$,  $a_i,  b_i, c_i \in \K[[X_1,\ldots,X_n]]$ satisfying Theorem~\ref{theorem:WPT}, with $b_0,\ldots,b_{d-1},c_0,\ldots,c_m$ known modulo ${\cal M}^k$, $\mathcal{M}$ the maximal ideal of $\K[[X_1,\ldots,X_n]]$.}
%		\newline
%		$\mathcal{F} = \{F_i\:\vert\:F_i = a_i - \sum_{j=0}^{i-1} b_jc_{i-j},\: i = 0,\ldots,d-1 \wedge i \leq m\}
%		\cup \{F_i\:\vert\:F_i = a_i - \sum_{j=0}^{m-1} \left(b_{i+j-m}\,c_{m-j}\right)\}$,
		\Ensure {$b_0,\ldots,b_{d-1},c_0,\ldots,c_m$ known modulo ${\cal M}^{k+1}$, updated in-place. }
		\vspace*{0.5em}
%		\Statex $\triangleright$ phase one
		\For{$i \gets 0$ to $d-1$} \Comment{phase 1}
		\State ${F_i}_{(k)} \gets {a_i}_{(k)}$
		\If {$i \leq m$}
			\For {$j \gets 0$ to $i-1$}
			\State ${F_i}_{(k)} \gets {F_i}_{(k)} - \left(b_j \, c_{i-j}\right)_{(k)}$
			\EndFor
		\Else
			\For {$j \gets 0$ to $m-1$}
			\State ${F_i}_{(k)} \gets {F_i}_{(k)} - \left(b_{i+j-m} \, c_{m-j}\right)_{(k)}$
			\EndFor
		\EndIf
		\State $s \gets 0$
		\For{$j \gets 1$ to $k-1$}
		\State $s \gets s\ +\ {b_i}_{(k - j)} \times {c_0}_{(j)}$
		\EndFor
		\State ${b_i}_{(k)}$ $\gets$ $\left( {F_i}_{(k)} - s\right) / {c_0}_{(0)}$
		\EndFor
		\vspace*{0.2em}
		\State ${c_m}_{(k)} \gets {a_{d+m}}_{(k)}$ \Comment{phase 2}
		\For{$i \gets 1$ to $m$}
			\If {$i \leq d$}
				\State ${c_{m-i}}_{(k)} \gets {a_{d+m-i}}_{(k)} - \sum_{j=1}^i \left(b_{d-j}c_{m-i+j}\right)_{(k)}$
			\Else
				\State ${c_{m-i}}_{(k)} \gets {a_{d+m-i}}_{(k)} - \sum_{j=1}^d \left(b_{d-j}c_{m-i+j}\right)_{(k)}$
			\EndIf
		\EndFor

	\end{algorithmic}
\end{algorithm}
}

By rearranging the first $d$ equations of (\ref{eqn:weier-eqs-upops})
and applying Lemma~\ref{lemma:l4WPT} we obtain ``phase 1'' of
\textsc{WeierstrassUpdate}, where each coefficient of $p$ is updated.
By rearranging the next $m+1$ equations of (\ref{eqn:weier-eqs-upops})
we obtain ``phase 2'' of \textsc{WeierstrassUpdate}, where each coefficient
of $\alpha$ is updated.
From Algorithm~\ref{alg:weierstrass-update}, it is then routine
to show the following two observations, which
lead to Theorem~\ref{theorem:weierstrass-complexity}.

\begin{Observation}[Weierstrass phase 1 complexity]
\label{obs:weierstrass-phase1}
$\ $For \textsc{Weierstrass\-Update} over $\K[[X_1]][Y]$, computing
${b_i}_{(k)}$, for $0 \leq i < d$, requires $2ki + 2k - 1$ operations in $\K$ if $i \leq m$, or $2km + 2k - 1$ operations in $\K$ if $i > m$.
\end{Observation}
\iflongversion
\myproof
Over $\K[[X_1]][Y]$, the homogeneous part of a coefficient is simply an element of $\K$.
Computing $s$ (Lines 9--11) requires $k-1$ multiplications and $k-2$ additions in $\K$.
Computing $(b_jc_{i-j})_{(k)}$ requires $2k-1$ operations in $\K$ (Observation~\ref{obs:ps-mult-ops}).
For $i \leq m$ there are $i$ such product homogeneous parts computed and $i$ subtractions in $\K$ to compute ${F_i}_{(k)}$
For $i > m$ there are $m$ such product homogeneous parts computed and $m$ subtractions in $\K$ to compute ${F_i}_{(k)}$.
Finally, 2 operations in $\K$ are required to compute ${b_i}_{(k)}$ from ${F_i}_{(k)}$, $s$, and ${c_0}_{(0)}$.
Hence, for $i < m$ the total is $2ki + 2k -1$, for $i \geq m$ the total is $2km + 2k - 1$.
\myfoorp
\fi

\begin{Observation}[Weierstrass phase 2 complexity]
\label{obs:weierstrass-phase2}
$\ $For \textsc{Weierstrass\-Update} over $\K[[X_1]][Y]$, computing
${c_{m-i}}_{(k)}$, for $0 \leq i < m$, requires $2ki$ operations in $\K$ if $i \leq d$, or $2kd$ operations in $\K$ if $i > d$.
\end{Observation}
\iflongversion
\myproof
Computing $\left(b_{d-j}c_{m-i+j}\right)_{(k)}$ requires $2k-1$ operations in $\K$
per Observation~\ref{obs:ps-mult-ops}.
For $i \leq d$ there are $i$ such product homogeneous parts and $i-1$ additions in $\K$.
For $i > d$ there are $d$ such product homogeneous parts and $d-1$ additions in $\K$.
Computing ${a_{d+m-1}}_{(k)}$ has no cost, since it is the input. Finally,
one subtraction in $\K$ finishes the computation of ${c_{m-i}}_{(k)}$.
Hence, for $i \leq d$ the total is $2ki$, for $i > d$ the total is $2kd$.
\myfoorp
\fi

\begin{Theorem}[Weierstrass preparation complexity]
\label{theorem:weierstrass-complexity}
Weierstrass preparation producing
$f=p\,\alpha$, with $f,p,\alpha \in \K[[X_1]][Y]$, $\deg{p} = d$, $\deg{\alpha}=m$,
%computing $p$ and $\alpha$ up to precision $k$ requires
requires $dmk^2 + dk^2 + dmk$ operations in $\K$
to compute $p$ and $\alpha$ to precision $k$.
\end{Theorem}
\myproof
Let $i$ be the index of a coefficient of $p$ or $\alpha$.
Consider the cost of computing the homogeneous part of degree $k$
of each coefficient of $p$ and $\alpha$.
First consider $i < t = \min(d,m)$. %that is, $b_0,\ldots,b_{t},c_0,\ldots,c_t$.
From Observations~\ref{obs:weierstrass-phase1} and \ref{obs:weierstrass-phase2}, computing
the $k$th homogeneous part of each $b_i$ and $c_i$ respectively requires
$2ki + 2k - 1$  and $2ki$ operations in $\K$.
For $0 \leq i < t$, this yields
a total of $2kt^2 + 2kt - t$.
Next, we have three cases: ($a$) $t=d=m$, ($b$) $m = t < i < d$, or ($c$) $d = t < i < m$.
In case ($a$) there is no additional work.
%and we have
%the total number of operations in $\K$ as $2dmk + 2kd -d$.
In case ($b$), phase 1 contributes an additional $(d-m)(2km + 2k - 1)$ operations.
%and phase 2 contributes no additional operations.
%In this case, $t = m$. Hence, the total number of operations in $\K$ is
%$2dmk+2dk-d$.
In case ($c$), phase 2 contributes an additional $(m-d)(2kd)$ operations.
%and phase 1 contributes no additional operations.
%In this case, $t=d$. Hence, the total number of operations in $\K$ is also
%$2dmk+2dk-d$.
In all cases, the total number of operations
to update $p$ and $\alpha$ from precision $k-1$ to precision $k$ is
$2dmk + 2dk -d$.
Finally, to compute $p$ and $\alpha$ up to precision $k$ requires
$dmk^2 + dk^2 + dmk$ operations in $\K$.
\myfoorp

%%%%%%%%%%%%%%%%
\iflongversion
\else
%%%%%%%%%%%%%%%%

A useful consideration is when the input to Weierstrass preparation is monic;
this arises for each application of WPT in Hensel factorization.
Then, $\alpha$ is necessarily monic, and the overall complexity of
Weierstrass preparation is reduced.
In particular, we save
computing $(b_{i-m}c_m)_{(k)}$
for the update of $b_i$, $i \geq m$ (Algorithm~\ref{alg:weierstrass-update}, Line 8),
and save computing $(b_{d-i}c_{m})_{(k)}$ for the
update of each $c_{m-i}$, $i \leq d$ (Algorithm~\ref{alg:weierstrass-update}, Line 16).
The following corollary states this result.

\begin{Corollary}[Weierstrass preparation complexity for monic input]$\,$
	\label{coro:wpt-monic}
	Weierstrass preparation producing
	$f=p\,\alpha$ with $f,p,\alpha \in \K[[X_1]][Y]$, $f$ monic in $Y$, $\deg{p} = d$
	and  $\deg{\alpha}=m$, requires $dmk^2 + dmk$ operations in $\K$
	to compute $p$ and $\alpha$ up to precision $k$.
\end{Corollary}

%%%%%%%%%%%%%%%%
\fi
%%%%%%%%%%%%%%%%

\subsection{Hensel Factorization}
\label{sec:hensel-complexity}

Before we begin Hensel factorization,
we will first see how to perform a translation,
or Taylor shift, by lazy evaluation.
For $f = \sum_{i=0}^d a_iY^i \in \K[[X_1,\ldots,X_n]][Y]$
and $c \in \K$, computing $f(Y+c)$ begins by pre-computing
the coefficients of the binomial expansions $(Y+c)^j$ for $0 \leq j \leq d$.
These coefficients are stored in a matrix $\mathbf{S}$.
Then, each coefficient of $f(Y+c) = \sum_{i=0}^d b_iY^i$ is a
linear combination of the coefficients of $f$
scaled by the appropriate elements of $\mathbf{S}$.
Since those elements of $\mathbf{S}$ are only elements of $\K$,
this linear combination does not change the degree % with respect to $X_1,\ldots,X_n$
and, for some integer $k$, ${b_i}_{(k)}$ relies only on
${a_\ell}_{(k)}$ for $i \leq \ell \leq d$.
This method is described in Algorithm~\ref{alg:lincombgen};
and its complexity is easily stated as Observation~\ref{obs:taylor-shift}.

\begin{algorithm}[t]
	\begin{algorithmic}[1]
		\algnotext{EndFor}
		\algorithmicfontsize
		\caption{{\sc TaylorShiftUpdate}($k$, $f$, $\mathbf{S}$, $i$)}\label{alg:lincombgen}
		\Require{For $f = \sum_{j=0}^d a_jY^j$, $g = f(Y+c) = \sum_{j=0}^d b_jY^j$, obtain the homogeneous part of degree $k$ for $b_i$.
		$\mathbf{S} \in \K^{(d+1) \times (d+1)}$ is a lower triangular matrix of coefficients of $(Y+c)^j$ for $j=0,\ldots,d$,}
		\Ensure {${b_i}_{(k)}$, the homogeneous part of degree $k$ of $b_i$.}
		\State $b_{i_{(k)}} \gets 0$
		\For{$\ell \gets i$ to $d$}
		\State $j \gets \ell + 1 - i$
		\State ${b_i}_{(k)} \gets {b_i}_{(k)} + S_{\ell+1,j} \times$ ${a_\ell}_{(k)}$
		\EndFor
		\State  \Return ${b_i}_{(k)}$
	\end{algorithmic}
\end{algorithm}
\begin{algorithm}[t!]
	\begin{algorithmic}[1]
		\algorithmicfontsize
		\caption{\small{\sc HenselFactorization}($f$)}\label{alg:hensel-factor}
		\algnotext{EndFor}
		\Require{$f = Y^d + \sum_{i=0}^{d-1} a_iY^i, a_i \in \K[[X_1,\ldots,X_n]]$.}
		\Ensure {$f_1,\ldots,f_r$ satisfying Theorem~\ref{theorem:hensel}.}
		\State $\bar{f} = f(0,\ldots,0,Y)$
		\State $(c_1,\ldots,c_r),(d_1,\ldots,d_r)  \gets$ roots and their multiplicities of $\bar{f}$ % \Comment{by some appropriate factorization algorithm}
		\State $c_1,\ldots,c_r \gets $ \textsc{sort}($[c_1,\ldots,c_r]$) by increasing multiplicity \Comment{see Theorem~\ref{theorem:hensel-per-factor}}
		\State $\hat{f}_1 \gets f$
		\For{$i \gets 1$ to $r-1$}
		\State $g_i \gets \hat{f}_i(Y+c_i)$
		\State $p_i, \alpha_i \gets $ \textsc{WeierstrassPreparation}($g$)
		\State $f_i \gets p_i(Y-c_i)$
		\State $\hat{f}_{i+1} \gets \alpha_i(Y-c_i)$
		\EndFor
		\State $f_{r} \gets \hat{f}_r$
		\State \Return $f_1,\ldots,f_r$
	\end{algorithmic}
\end{algorithm}
\begin{figure}[t!]
	\centering

	\begin{tikzpicture}[node distance=1.25cm, font={\normalsize}]
	\node[] (f) at (0,0) {$f$};
	\node[right of=f] (g1) {$g_1$};
	\node[right of=g1] (a1) {$\alpha_1$};
	\node[above=0.5cm of a1] (p1) {$p_1$};
	\node[right of=p1] (f1) {$f_1$};
	\node[right of=a1] (fhat1) {$\hat{f}_2$};
	\node[right of=fhat1] (g2) {$g_2$};

	\node[right of=g2] (a2) {$\alpha_2$};
	\node[above=0.5cm of a2] (p2) {$p_2$};
	\node[right of=p2] (f2) {$f_2$};
	\node[right of=a2] (fhat2) {$\hat{f}_3$};
	\node[right of=fhat2] (g3) {$g_3$};

	\node[right of=g3] (a3) {$\alpha_3$};
	\node[above=0.5cm of a3] (p3) {$p_3$};
	\node[right of=p3] (f3) {$f_3$};
	\node[right of=a3] (f4) {$f_4$};

	\draw[blue, ->, >={Stealth}] (f) -- node[above,midway,font=\scriptsize] {$+c_1$} (g1);
	\draw[blue, ->, >={Stealth}] (g1) -- (p1);
	\draw[blue, ->, >={Stealth}] (g1) -- (a1);
	\draw[blue, ->, >={Stealth}] (p1) -- node[above,midway,font=\scriptsize] {$-c_1$} (f1);

	\draw[red!70!black, ->, >={Stealth}] (a1) -- node[above,midway,font=\scriptsize] {$-c_1$} (fhat1);
	\draw[red!70!black, ->, >={Stealth}] (fhat1) -- node[above,midway,font=\scriptsize] {$+c_2$} (g2);
	\draw[red!70!black, ->, >={Stealth}] (g2) -- (a2);
	\draw[red!70!black, ->, >={Stealth}] (g2) -- (p2);
	\draw[red!70!black, ->, >={Stealth}] (p2) -- node[above,midway,font=\scriptsize] {$-c_2$} (f2);

	\draw[green!50!black, ->, >={Stealth}] (a2) -- node[above,midway,font=\scriptsize] {$-c_2$} (fhat2);
	\draw[green!50!black, ->, >={Stealth}] (fhat2) -- node[above,midway,font=\scriptsize] {$+c_3$} (g3);
	\draw[green!50!black, ->, >={Stealth}] (g3) -- (p3);
	\draw[green!50!black, ->, >={Stealth}] (p3) -- node[above,midway,font=\scriptsize] {$-c_3$} (f3);
	\draw[green!50!black, ->, >={Stealth}] (g3) -- (a3);
	\draw[frenchplum!80!blue, ->, >={Stealth}] (a3) -- node[above,midway,font=\scriptsize] {$-c_3$} (f4);
	\end{tikzpicture}
	\caption{The ancestor chain for the Hensel factorization $f = f_1f_2f_3f_4$.
		Updating $f_1$ requires updating $g_1, p_1, \alpha_1$; then
		updating $f_2$ requires updating $\hat{f}_2, g_2, p_2, \alpha_2$;
		then updating $f_3$ requires updating $\hat{f}_3, g_3, p_3, \alpha_3$;
		then updating $f_4$ requires only its own Taylor shift.
		These groupings form the eventual stages of the Hensel pipeline (Algorithm~\ref{alg:hensel-factor-pipeline}).}\label{fig:hensel-pipeline}
\end{figure}
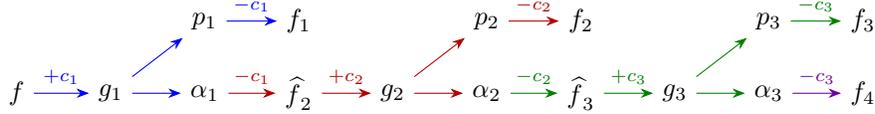

\begin{Observation}[Taylor shift complexity]
	\label{obs:taylor-shift}
	For a UPoPS $f = \sum_{i=0}^d a_iY^i \in \K[[X_1]][Y]$, computing the homogeneous part of degree $k$ for all coefficients of the shifted UPoPS $f(Y+c)$
	requires $d^2 + 2d + 1$ operations in $\K$.
\end{Observation}
\iflongversion
\myproof
Computing $f(Y+c)$ requires updating $d+1$ power series coefficients via \textsc{TaylorShiftUpdate}.
Computing the homogeneous part of degree $k$ of the $i$th coefficient of $f(Y+c)$ requires $2d - 2i + 1$ operations in $\K$:
$d-i+1$ multiplications and $d-i$ additions.
Summing over $i$ from $0$ to $d$ yields $d^2 + 2d + 1$.
\myfoorp
\fi

Having specified the update functions for WPT and Taylor shift,
lazy Hensel factorization is immediate,
requiring only the appropriate chain of ancestors.
Algorithm~\ref{alg:hensel-factor} shows this initialization
through repeated applications
of Taylor shift and Weierstrass preparation.
Note that factors are sorted by increasing degree
to enable better load-balance in the eventual parallel algorithm.
Fig.~\ref{fig:hensel-pipeline} shows
the chain of ancestors created by $f = f_1f_2f_3f_4$
and the grouping of ancestors required to update each factor; the complexity of
which is shown in Theorem~\ref{theorem:hensel-per-factor}.
Corollary~\ref{coro:hensel-per-iter} follows immediately
and Corollary~\ref{coro:hensel-general-roots} gives
the total complexity of Hensel factorization.
Here, we ignore the
%% little
initial cost of factorizing $\bar{f}$.

%%%%%%%%%%%%%
%%%%%%%%%%%%%

\begin{Theorem}[Hensel factorization complexity per factor]
\label{theorem:hensel-per-factor}
Let ${\hat{d}_i}$ be the degree of $\hat{f}_i$ during
\textsc{HenselFactorization} applied to $f \in \K[[X_1]][Y]$, $\deg{f} = d$.
To update $f_1$, $\deg{f_1} = d_1$ to precision $k$ requires
$d_1{\hat{d}_2}k^2 + {d}^2k + d_1dk + 2d_1k + 2dk + 2k$ operations in $\K$.
To update $f_i$, $\deg{f_i} = d_i$, for $1 < i < r$, to precision $k$ requires
$d_i{\hat{d}_{i+1}}k^2 + 2{\hat{d}_i}^2k + d_i{\hat{d}_i}k + 2d_ik + 4{\hat{d}_i}k + 3k$
operations in $\K$.
To update $f_r$, $\deg{f_r} = d_r$, to precision $k$
requires $d_r^2k + 2d_rk + k$ operations in $\K$.
\end{Theorem}
\myproof
Updating the first factor produced by \textsc{HenselFactorization}
requires one Taylor shift of degree $d$, one Weierstrass
preparation producing $p_1$ and $\alpha_1$ of degree $d_1$ and ${\hat{d}_2} = d-d_1$,
and one Taylor shift of degree $d_1$ to obtain $f_1$ from $p$.
From Observation~\ref{obs:taylor-shift} and Corollary~\ref{coro:wpt-monic}
we have that the Taylor shifts require $k(d^2 + 2d +1) + k(d_1^2 + 2d_1 + 1)$
operations in $\K$ and the Weierstrass preparation requires
$d_1(d-d_1)k^2 + d_1(d-d_1)k$ operations in $\K$.
The total cost counted as operations in $\K$ is thus
$d_1{\hat{d}_2}k^2 + {d}^2k + d_1dk + 2d_1k + 2dk + 2k$.

Updating each following factor, besides the last, requires
one Taylor shift of degree $\hat{d}_i$ to update $\hat{f}_i$ from
$\alpha_{i-1}$, one Taylor shift of degree $\hat{d}_i$ to update $g_i$ from
$\hat{f}_i$, one Weierstrass preparation to obtain $p_i$ and $\alpha_i$
of degree $d_i$ and $\hat{d}_{i+1} = \hat{d}_{i} - d_i$, and
one Taylor shift of degree $d_i$ to obtain $f_i$ from $p_i$.
The Taylor shifts require $2k({\hat{d}_i}^2 + 2{\hat{d}_i} +1) + k(d_i^2 + 2d_i + 1)$ operations in $\K$.
The Weierstrass preparation requires $d_i(\hat{d}_{i}-d_i)k^2 + d_i(\hat{d}_{i}-d_i)k$ operations in $\K$.
The total cost counted as operations in $\K$ is thus $d_i{\hat{d}_{i+1}}k^2 + 2{\hat{d}_i}^2k + d_i{\hat{d}_i}k + 2d_ik + 4{\hat{d}_i}k + 3k.$

Finally, updating the last factor to precision $k$ requires a single Taylor shift of degree $d_r$ costing $d_r^2k + 2d_rk + k$ operations in $\K$.
\myfoorp

\begin{Corollary}[Hensel factorization complexity per iteration]
\label{coro:hensel-per-iter}
Let $\hat{d}_i$ be the degree of $\hat{f}_i$ during the
\textsc{HenselFactorization} algorithm applied to $f \in \K[[X_1]][Y]$, $\deg{f} = d$.
Computing the $k$th homogeneous part of $f_1$, $\deg{f_1} = d_1$, requires
$2d_1\hat{d}_2k + d_1^2 + d^2 + 2d_1 + 2d + 2$ operations in $\K$.
Computing the $k$th homogeneous part of $f_i$, $\deg{f_i} = d_i$, $1 < i < r$, requires
$2d_i\hat{d}_{i+1}k + d_i^2 + 2{\hat{d}_i}^2 + 4\hat{d}_i + 2d_i + 3$ operations in $\K$.
Computing the $k$th homogeneous part of $f_r$, $\deg{f_r} = d_r$,
requires $d_r^2 + 2d_r + 1$ operations in $\K$.
\end{Corollary}
\iflongversion
\myproof
Follows from Observation~\ref{obs:taylor-shift},
Corollary~\ref{coro:wpt-monic}, and Theorem~\ref{theorem:hensel-per-factor}.
\myfoorp
\fi

\begin{Corollary}[Hensel factorization complexity]
\label{coro:hensel-general-roots}
\textsc{HenselFactorization} producing
$f = f_1\cdots f_r$, with $f \in \K[[X_1]][Y]$, $\deg{f} = d$,
requires $\mathcal{O}(d^3k + d^2k^2)$ operations in $\K$
to update all factors to precision $k$.
\end{Corollary}
\myproof
Let $f_1,\ldots,f_r$ have respective degrees $d_1,\ldots,d_r$.
Let $\hat{d_i} = \sum_{j=i}^r d_j$ (thus $\hat{d}_1 = d$ and $\hat{d}_r = d_r$).
From Theorem~\ref{theorem:hensel-per-factor}, each $f_i$,  $1 \leq i < r$
requires $\mathcal{O}(d_i\hat{d}_{i+1}k^2 + {\hat{d}_i}^2k)$
operations in $\K$ to be updated
to precision $k$ (or $\mathcal{O}(d_r^2k)$ for $f_r$).
We have $\sum_{i=1}^{r-1} d_i\hat{d}_{i+1} \leq \sum_{i=1}^{r-1} d_i d < d^2$
and $\sum_{i=1}^r {\hat{d}_i}^2 \leq \sum_{i=1}^r d^2 = rd^2 \leq d^3$.
Hence, all factors can be updated to precision $k$ within $\mathcal{O}(d^3k + d^2k^2)$
operations in $\K$.
\myfoorp

Corollary~\ref{coro:hensel-general-roots} shows that the two dominant terms
in the cost of computing a Hensel factorization of a UPoPS of degree $d$,
up to precision $k$, are $d^3k$ and $d^2k^2$.
From the proof of Theorem~\ref{theorem:hensel-per-factor},
the former term arises from the cost of the Taylor shifts in $Y$,
meanwhile, the latter term arises from the (polynomial)
multiplication of homogeneous parts in Weierstrass preparation.
This observation then leads to the following conjecture.
Recall that $M(n)$ denotes a polynomial multiplication
time \cite[Ch. 8]{Gathen:2003:MCA:945759}.
From \cite{DBLP:journals/jsc/Hoeven02a},
relaxed algorithms, which improve the performance
of lazy evaluation schemes,
can be used to compute a power series product
in $\K[[X_1]]$ up to precision $k$
in at most $\mathcal{O}(M(k)\log k)$ operations in $\K$
(or less, in view of the improved relaxed multiplication
of \cite{DBLP:conf/issac/Hoeven14}).

\begin{Conjecture}
\label{conjecture:relax}
Let $f \in \K[[X_1]][Y]$
factorize as $f_1\cdots f_r$ using \textsc{HenselFactorization}.
Let $\deg{f} = d$.
Updating the factors $f_1,\ldots,f_r$ to precision $k$
using relaxed algorithms
requires at most $\mathcal{O}(d^3k + d^2M(k)\log k)$
operations in $\K$.
\end{Conjecture}

Comparatively, the Hensel--Sasaki Construction
requires at most $\mathcal{O}(d^3M(d) + dM(d)k^2)$
operations in $\K$ to compute the first $k$ terms of all factors
of $f \in \K[X_1,Y]$, where $f$ has total degree $d$ \cite{DBLP:journals/jsc/AlvandiAKM20}.
The method of Kung and Traub \cite{DBLP:journals/jacm/KungT78},
requires
$\mathcal{O}(d^2M(k))$. % operations in $\K$.
Already, Corollary~\ref{coro:hensel-general-roots}---where $d = \deg{f,Y}$---shows
that our Hensel factorization is an improvement on Hensel--Sasaki
($d^2k^2$ versus $dM(d)k^2$). %by a factor of at least $\log k$.
If Conjecture~\ref{conjecture:relax} is true, then
Hensel factorization can be within a factor of $\log k$
of Kung and Traub's method.
Nonetheless, this conjecture is highly encouraging where $k \gg d$,
particularly where we have already seen that our current, suboptimal,
method performs better in practice than Hensel--Sasaki and
the method of Kung and Traub \cite{DBLP:conf/casc/BrandtKM20}.
Proving this conjecture is left to future work.

%% file: parallel.tex
\section{Parallel Algorithms}
\label{sec:parallel}

Section~\ref{sec:complexity}
presented lazy algorithms for Weierstrass preparation,
Taylor shift, and Hensel factorization.
It also presented complexity estimates for those algorithms.
Those estimates will soon be used to help dynamically
distribute hardware resources (threads)
in a parallel variation of Hensel factorization;
in particular, a Hensel factorization pipeline
where each pipeline stage updates one or more factors,
see Algorithms~\ref{alg:hensel-pipeline-stage}--\ref{alg:hensel-distrib}.
But first, we will examine parallel processing techniques
for Weierstrass preparation.

\subsection{Parallel Algorithms for Weierstrass Preparation}
\label{sec:para-weierstrass}

Algorithm~\ref{alg:weierstrass-update}
shows that $p$ and $\alpha$ from a Weierstrass preparation
can be updated in two phases: $p$ in phase 1, and $\alpha$ in phase 2.
Ultimately, these updates rely on the computation
of the homogeneous part of some power series product.
Algorithm~\ref{alg:update-deg-parallel} presents
a simple map-reduce pattern (see Section~\ref{sec:parallel-patterns})
for computing such a homogeneous part.
Moreover, this algorithm is designed such that, recursively,
all ancestors of a power series product are also updated
using parallelism. Note that \textsc{UpdateToDegParallel}
called on a UPoPS simply recurses on each of its coefficients.

\begin{algorithm}[t!]
	\begin{algorithmic}[1]
		\algnotext{EndFor}
		\algnotext{EndIf}
		\algnotext{EndParaFor}
		\algorithmicfontsize
		\caption{\small{\sc UpdateToDegParallel}($k$, $f$, $t$)}\label{alg:update-deg-parallel}
		\Require{A positive integer $k$, $f \in \K[[X_1,\ldots,X_n]]$ known to at least precision $k-1$. If $f$ has ancestors,
			it is the result of a binary operation. A positive integer $t$ for the number of threads to use.}
		\Ensure {$f$ is updated to precision $k$, in place.}
		\If {$f_{(k)}$ already computed}
		\State \Return
		\EndIf
		\State $g,\ h \gets$ \textsc{FirstAncestor}($f$), \textsc{SecondAncestor}($f$)
		\State \textsc{UpdateToDegParallel}($k$, $g$, $t$);
		\State \textsc{UpdateToDegParallel}($k$, $h$, $t$);
		\If {$f$ is a product}
		\State $\mathcal{V} \gets [0,\ldots,0]$ \Comment{0-indexed list of size $t$}
		\ParaFor {$j \gets 0$ to $t-1$}
		\For {$i \gets \nicefrac{jk}{t}$ to $\nicefrac{(j+1)k}{t} - 1$ \textbf{while} $i \leq k$}
		\State $\mathcal{V}[j] \gets \mathcal{V}[j] + g_{(i)}h_{(k-i)}$
		\EndFor
		\EndParaFor
		\State $f_{(k)} \gets \sum_{j=0}^{t-1} \mathcal{V}[j]$ \Comment{reduce}
		\ElsIf {$f$ is a $p$ from a Weierstrass preparation}
		\State \textsc{WeierstrassPhase1Parallel}($k$,$g$,$f$,$h$,\textsc{WeierstrassData}($f$),$t$)
		\ElsIf {$f$ is an $\alpha$ from a Weierstrass preparation}
		\State \textsc{WeierstrassPhase2Parallel}($k$, $g$, $h$, $f$, $t$)
		\Else
		\State \textsc{UpdateToDeg}($k$, $f$)
		\EndIf
	\end{algorithmic}
\end{algorithm}

Using the notation of Algorithm~\ref{alg:weierstrass-update},
recall that, e.g.,
${F_i} \gets {a_i} - \sum_{j=0}^{i-1} (b_jc_{i-j})$,
for $i \leq m$.
Using lazy power series arithmetic, this entire formula
can be encoded by a chain of ancestors, and one
simply needs to update ${F_i}$ to trigger
a cascade of updates through its ancestors.
In particular, using Algorithm~\ref{alg:update-deg-parallel},
the homogeneous part of each product $b_jc_{i-j}$
is recursively computed using map-reduce.
Similarly, Lemma~\ref{lemma:l4WPT}
can be implemented using map-reduce (see Algorithm~\ref{alg:lemma-para})
to replace Lines 9--12 of Algorithm~\ref{alg:weierstrass-update}.
Phase 1 of Weierstrass, say \textsc{Weierstrass\-Phase1\-Parallel},
thus reduces to a loop over $i$ from 0 to $d-1$, calling Algorithm~\ref{alg:update-deg-parallel}
to update $F_i$ to precision $k$, and calling Algorithm~\ref{alg:lemma-para}
to compute ${b_i}_{(k)}$.
%This parallel phase 1 update is summarized as Algorithm~\ref{alg:weierstrass-para-update1}.

Algorithm~\ref{alg:update-deg-parallel} uses several simple subroutines:
\textsc{FirstAncestor} and \textsc{SecondAncestor} gets the
first and second ancestor of a power series,
\textsc{WeierstrassData} gets a reference to the list of $F_i$'s, and
\textsc{UpdateToDeg} calls the serial update function of a lazy power series
to ensure its precision is at least $k$; see Section~\ref{sec:lazy-ps}.

\begin{algorithm}[t!]
	\begin{algorithmic}[1]
		\algnotext{EndFor}
		\algnotext{EndIf}
		\algorithmicfontsize
		\caption{\small{\sc LemmaForWeierstrass}($k$, $f$, $g$, $h$, $t$)}\label{alg:lemma-para}
		\Require{$f,g,h \in \K[[X_1,\ldots,X_n]]$ such that $f=gh$, $f_{(0)} = 0$, $h_{(0)} \neq 0$, $f$ known to precision $k$, and $g,h$ known to precision $k-1$. $t \geq 1$ the number of threads to use.}
		%		\newline
		%		$\mathcal{F} = \{F_i\:\vert\:F_i = a_i - \sum_{j=0}^{i-1} b_jc_{i-j},\: i = 0,\ldots,d-1 \wedge i \leq m\}
		%		\cup \{F_i\:\vert\:F_i = a_i - \sum_{j=0}^{m-1} \left(b_{i+j-m}\,c_{m-j}\right)\}$,
		\Ensure {$g_{(k)}$.}
		\State $\mathcal{V} \gets [0,\ldots,0]$ \Comment{0-indexed list of size $t$}
		\ParaFor {$j \gets 0$ to $t-1$}
		\For {$i \gets \nicefrac{jk}{t} + 1$ to $\nicefrac{(j+1)k}{t}$ \textbf{while} $i < k$}
		\State $\mathcal{V}[j] \gets \mathcal{V}[j] + g_{(k-i)}h_{(i)}$
		\EndFor
		\EndParaFor
		\State \Return $\left( {f}_{(k)} - \sum_{j=0}^{t-1} \mathcal{V}[j]\right) / {h}_{(0)}$
	\end{algorithmic}
\end{algorithm}

\iffalse
\begin{algorithm}[tb]
	\begin{algorithmic}[1]
		\algnotext{EndFor}
		\algnotext{EndIf}
		\algorithmicfontsize
		\caption{\small{\sc WeierstrassPhase1Parallel}($k$, $f$, $p$, $\alpha$, $\mathcal{F}$, $t$)}\label{alg:weierstrass-para-update1}
		\Require{$f = \sum_{i=0}^{d+m} a_iY^i$, $p = Y^d + \sum_{i=0}^{d-1} b_iY^i$, $\alpha = \sum_{i=0}^m c_iY^i$,  $a_i,  b_i, c_i \in \K[[X_1,\ldots,X_n]]$ satisfying Theorem~\ref{theorem:WPT}.	$b_0,\ldots,b_{d-1},c_0,\ldots,c_m$ known modulo ${\cal M}^k$, $\mathcal{M}$ the maximal ideal of $\K[[X_1,\ldots,X_n]]$.
		$\mathcal{F} \gets \{F_i\:\vert\:F_i = a_i - \sum_{j=0}^{i-1} b_jc_{i-j},\, i \leq d-1 \,\wedge\, i \leq m\}
		\cup \{F_i\:\vert\:F_i = a_i - \sum_{j=0}^{m-1} b_{i+j-m}\,c_{m-j},\, i \leq d-1 \,\wedge\, i > m\}$.  A positive integer $t$ for the number of threads to use.}
		\Ensure {$b_0,\ldots,b_{d-1}$ known modulo ${\cal M}^{k+1}$, updated in-place.}
		\For{$i \gets 0$ to $d-1$}
		\State \textsc{UpdateToDegParallel}($k$, $F_i$, $t$)
		\State ${b_i}_{(k)}$ $\gets$ \textsc{LemmaForWeierstrass}($k$, $F_i$, $b_i$, $c_0$, $t$)
		\EndFor
	\end{algorithmic}
\end{algorithm}
\fi

%\setlength\textfloatsep{10pt plus 2pt minus 2pt}
\begin{algorithm}[t!]
	\begin{algorithmic}[1]
		\algnotext{EndFor}
		\algnotext{EndIf}
		\algnotext{EndParaFor}
		\algorithmicfontsize
		\caption{\small{\sc WeierstrassPhase2Parallel}($k$, $f$, $p$, $\alpha$, $t$)}\label{alg:weierstrass-para-update2}
		\Require{$f = \sum_{i=0}^{d+m} a_iY^i$, $p = Y^d + \sum_{i=0}^{d-1} b_iY^i$, $\alpha = \sum_{i=0}^m c_iY^i$,  $a_i,  b_i, c_i \in \K[[X_1,\ldots,X_n]]$ satisfying Theorem~\ref{theorem:WPT}.	$b_0,\ldots,b_{d-1}$ known modulo ${\cal M}^{k+1}$, $c_0,\ldots,c_m$ known modulo ${\cal M}^k$, for $\mathcal{M}$ the maximal ideal of $\K[[X_1,\ldots,X_n]]$.
			$t \geq 1$ for the number of threads to use.}
		\Ensure {$c_0,\ldots,c_m$ known modulo ${\cal M}^{k+1}$, updated in-place.}
		\State $work \gets 0$
		\For{$i \gets 1$ to $m$} \Comment{estimate work using Observation~\ref{obs:weierstrass-phase2}, Corollary~\ref{coro:wpt-monic}}
		\If {$i \leq d$} $\ work \gets work + i - (a_{d+m} = 0)$ \Comment{eval. Boolean as an integer}
%		\ElsIf {$i \leq d$ \textbf{and} $a_{d+m} = 1$} $\ work \gets work + (i-1)$
		\Else $\ work\gets work + d$
		\EndIf
		\EndFor
		\State $t' \gets 1$; $targ \gets work \,/\, t$
		\State $work \gets 0$; $j \gets 1$
		\State $\mathcal{I} \gets [-1,0,\ldots,0]$ \Comment{0-indexed list of size $t+1$}
		\For{$i \gets 1$ to $m$}
		\If {$i \leq d$} $\ work \gets work + i - (a_{d+m} = 0)$
%		\ElsIf {$i \leq d$ \textbf{and} $a_{d+m} = 1$} $\ work \gets work + (i-1)$
		\Else $\ work \gets work + d$
		\EndIf
		\If {$work \geq targ$}
		\State $\mathcal{I}[j] \gets i$; $work \gets 0$; $j \gets j + 1$
		\EndIf
		\EndFor
		\If {$j \leq t$ \textbf{and} $t' < 2$} \Comment{work did not distribute evenly; try again with $t=\nicefrac{t}{2}$}
		\State $t \gets t \,/\, 2$; $t' \gets 2$
		\State \textbf{goto} Line~6
		\ElsIf {$j \leq t$} \Comment{still not even, use all threads in \textsc{UpdateToDegParallel}}
		\State $\mathcal{I}[1] \gets m$; $t' \gets 2t$; $t \gets 1$
		\EndIf
		\ParaFor {$\ell \gets 1$ to $t$}
		\For {$i \gets \mathcal{I}[\ell-1]+1$ to $\mathcal{I}[\ell]$}
		\State \textsc{UpdateToDegParallel}($k$, $c_{m-i}$, $t'$)
		\EndFor
		\EndParaFor
		%		\If {$t > 1$}
		%		\ParaFor {$\ell = 1$ to $t$}
		%			\For {$i = \mathcal{I}[\ell-1]+1$ to $\mathcal{I}[\ell]$}
		%				\State \textsc{UpdateToDegParallel}($k$, $c_{m-i}$, $t'$)
		%			\EndFor
		%		\EndParaFor
		%		\Else
		%			\For {$i = 0$ to $m$}
		%				\State \textsc{UpdateToDegParallel}($k$, $c_{m-i}$, $t'$)
		%			\EndFor
		%		\EndIf
	\end{algorithmic}
\end{algorithm}

Now consider phase 2 of \textsc{WeierstrassUpdate}.
Notice that computing the homogeneous part of degree $k$
for $c_{m-i}$, $0 \leq i \leq m$ only requires
each $c_{m-i}$ to be known up to precision $k-1$,
since each $b_j \in \mathcal{M}$ for $0 \leq j < d$.
This implies that the phase 2 \textbf{for} loop of
\textsc{WeierstrassUpdate} has independent iterations.
We thus apply the map pattern directly to this loop itself,
rather than relying on the map-reduce pattern of \textsc{UpdateToDegParallel}.
However, consider the following two facts:
the cost of computing each $c_{m-i}$ is different
(Observation~\ref{obs:weierstrass-phase2} and Corollary~\ref{coro:wpt-monic}),
and, for a certain number of available threads $t$,
it may be impossible to partition the iterations
of the loop into $t$ partitions of equal work.
Yet, partitioning the loop itself
is preferred for greater parallelism.

Hence, for phase 2, a dynamic decision
is made to either apply the map pattern to
the loop over $c_{m-i}$, or to apply the map pattern
within \textsc{UpdateToDegParallel} for each $c_{m-i}$,
or both.
This decision process is detailed in Algorithm~\ref{alg:weierstrass-para-update2},
where $t$ partitions of equal work try to be found
to apply the map pattern to only the loop itself.
If unsuccessful, $\nicefrac{t}{2}$ partitions of equal work
try to be found, with 2 threads to be used within
\textsc{UpdateToDegParallel} of each partition. If that, too, is unsuccessful,
then each $c_{m-i}$ is updated one at a time using
the total number of threads $t$ within \textsc{UpdateToDegParallel}.

\subsection{Parallel Algorithms for Hensel Factorization}
\label{sec:para-hensel}

Let $f = f_1\cdots f_r$ be a Hensel factorization
where the factors have respective degrees $d_1,\ldots,d_r$.
From Algorithm~\ref{alg:hensel-factor}
and Figure~\ref{fig:hensel-pipeline},
we have already seen that the repeated applications of
Taylor shift and Weierstrass preparation naturally
form a chain of ancestors, and thus a pipeline.
Using the notation of Algorithm~\ref{alg:hensel-factor},
updating $f_1$ requires updating
$g_1,p_1,\alpha_1$. Then, updating $f_2$ requires
updating $\hat{f}_2, g_2, p_2, \alpha_2$, and so on.
These groups easily form stages of a pipeline,
where updating $f_1$ to degree $k-1$ is a prerequisite
for updating $f_2$ to degree $k-1$.
Moreover, meanwhile $f_2$ is being updated to degree $k-1$,
$f_1$ can simultaneously be updated to degree $k$.
Such a pattern holds for all successive factors.

{
	%\begin{figure}[tb]
	%\setlength\floatsep{0.2em}
	\begin{algorithm}[t!]
		\begin{algorithmic}[1]
			\algorithmicfontsize
			\caption{\small{\sc HenselPipelineStage}($k$, $f_i$, $t$, \textsc{gen})}\label{alg:hensel-pipeline-stage}
			\algnotext{EndFor}
			\Require{A positive integer $k$, $f_i = Y^{d_i} + \sum_{i=0}^{{d_i}-1} a_iY^i, a_i \in \K[[X_1,\ldots,X_n]]$. A positive integer $t$ the number of threads to use within this stage. \textsc{gen} a generator for the previous pipeline stage.}
			\Ensure {a sequence of integers $j$ signalling $f_i$ is known to precision $j$. This sequence ends with $k$.}
			\State $p \gets$ \textsc{Precision}($f_i$) \Comment{get the current precision of $f_i$}
			\Do
			\State {$k' \gets$  \textsc{gen}()} \Comment{A blocking function call until \textsc{gen} yields}
			\For {$j \gets p$ to $k'$}
			\State \textsc{UpdateToDegParallel}($j$, $f_i$, $t$)
			\State \yield $\ j$
			\EndFor
			\State $p \gets k'$
			\doWhile {$k' < k$}
		\end{algorithmic}
	\end{algorithm}
	\begin{algorithm}[t!]
		\begin{algorithmic}[1]
			\algorithmicfontsize
			\caption{\small{\sc HenselFactorizationPipeline}($k$, $\mathcal{F}$, $\mathcal{T}$)}\label{alg:hensel-factor-pipeline}
			\algnotext{EndFor}
			\algnotext{EndIf}
			\Require{A positive integer $k$, $\mathcal{F} = \{f_1,\ldots,f_r\}$, the output of \textsc{HenselFactorization}. $\mathcal{T} \in \mathbb{Z}^r$ a 0-indexed list of the number of threads to use in each stage, $\mathcal{T}[r-1] > 0$.}
			\Ensure {$f_1,\ldots,f_r$ updated in-place to precision $k$.}
			\State \textsc{gen} $\gets\ (\,) \rightarrow $ $\{$\yield $\ k\}$ \Comment{An anonymous function asynchronous generator}
			\For {$i \gets 0$ to $r-1$}
			\If {$\mathcal{T}[i] > 0$}
			\Statex $\qquad\quad\triangleright$ Capture \textsc{HenselPipelineStage}($k, f_{i+1}, \mathcal{T}[i]$, \textsc{gen}) as a
			\Statex $\qquad\quad\ \,$ function object, passing the previous \textsc{gen} as input
			\State \textsc{gen} $\gets$ \textsc{AsyncGenerator}(\textsc{HenselPipelineStage}, $k$, $f_{i+1}$, $\mathcal{T}[i]$, \textsc{gen})
			\EndIf
			\EndFor
			\Do
			\State $k' \gets$ \textsc{gen}() \Comment{ensure last stage completes before returning}
			\doWhile {$k' < k$}
		\end{algorithmic}
	\end{algorithm}
	%\end{figure}

}

Algorithms~\ref{alg:hensel-pipeline-stage} and \ref{alg:hensel-factor-pipeline}
show how the factors of a Hensel factorization can all be simultaneously
updated to degree $k$ using asynchronous generators, denoted by the constructor \textsc{AsyncGenerator},
forming the so-called \textit{Hensel pipeline}.
Algorithm~\ref{alg:hensel-pipeline-stage} shows a single pipeline stage
as an asynchronous generator, which itself consumes data from another
asynchronous generator---just as expected from the pipeline pattern.
Algorithm~\ref{alg:hensel-factor-pipeline} shows the creation, and joining in sequence,
of those generators. The key feature of these algorithms
is that a generator (say, stage $i$) produces a sequence of integers $(j)$ which signals
to the consumer (stage $i+1$) that the previous factor
has been computed up to precision $j$ and thus
the required data is available to update its own factor to precision $j$.

Notice that Algorithm~\ref{alg:hensel-factor-pipeline}
still follows our lazy evaluation scheme.
Indeed, the factors are updated all at once up to precision $k$,
starting from their current precision.
However, for optimal performance, the updates should be applied
for large increases in precision,
rather than repeatedly increasing precision by one.

Further considering performance, Theorem~\ref{theorem:hensel-per-factor}
showed that the cost for updating
each factor of a Hensel factorization
is different.
In particular, for $\hat{d}_i \gets \sum_{j=i}^r d_j$,
updating factor $f_i$ scales as $d_i\hat{d}_{i+1}k^2$.
The work for each stage of the proposed pipeline
is unequal and the pipeline is unlikely to achieve good parallel speedup.
However, Corollary~\ref{coro:hensel-per-iter}
shows that the work ratios between stages do not change for increasing $k$,
and thus a static scheduling scheme is sufficient.
%Further, Corollary~\ref{coro:hensel-per-iter} shows
%that the work ratios between each factor does not change between iterations
%or increasing precision.

Notice that Algorithm~\ref{alg:hensel-pipeline-stage}
takes a parameter $t$ for the number of threads to use internally.
As we have seen in Section~\ref{sec:para-weierstrass},
the Weierstrass update can be performed in parallel.
Consequently, each stage of the Hensel pipeline uses $t$ threads
to exploit such parallelism.
We have thus composed the two parallel schemes,
applying map-reduce within each stage of the parallel pipeline.
This composition serves to load-balance the pipeline.
For example, the first stage may be given $t_1$ threads
and the second stage given $t_2$ threads, with $t_1 > t_2$,
so that the two stages may execute in nearly equal time.

To further encourage load-balancing, each stage of the pipeline
need not update a single factor, but rather a group of successive factors.
Algorithm~\ref{alg:hensel-distrib} applies Theorem~\ref{theorem:hensel-per-factor}
to attempt to load-balance each stage $s$ of the pipeline
by assigning a certain number of threads $t_s$
and a certain group of factors $f_{s_1},\ldots,f_{s_2}$ to it.
The goal is for $\sum_{i=s_1}^{s_2} d_i\hat{d}_{i+1} \,/\, t_s$
to be roughly equal for each stage.

\begin{algorithm}[tb]
	\begin{algorithmic}[1]
		\algorithmicfontsize
		\caption{\small{\sc DistributeResourcesHensel}($\mathcal{F}$, $t_{tot}$)}\label{alg:hensel-distrib}
		\algnotext{EndFor}
		\algnotext{EndIf}
		\Require{$\mathcal{F} = \{f_1,\ldots,f_r\}$ the output of \textsc{HenselFactorization}. $t_{tot} > 1$ the total number of threads.}
		\Ensure {$\mathcal{T}$, a list of size $r$, where $\mathcal{T}[i]$ is the number of threads to use for updating $f_{i+1}$.
			The number of positive entries in $\mathcal{T}$ determines the number of pipeline stages. $\mathcal{T}[i] = 0$ encodes that $f_{i+1}$ should be computed within the same stage as $f_{i+2}$.}
		\State $\mathcal{T} \gets [0,\ldots,0,1]$; $t \gets t_{tot} - 1$ \Comment{$\mathcal{T}[r-1] = 1$ ensures last factor gets updated}
%		\If {$t > 0$}
		\State $d \gets \sum_{i=1}^r \deg{f_i}$
		\State $\mathcal{W} \gets [0,\ldots,0]$ \Comment{A 0-indexed list of size $r$}
		\For {$i \gets 1$ to $r-1$}
		\State $\mathcal{W}[i-1] \gets \deg{f_i}(d - \deg{f_i})$ \Comment{Estimate work by Theorem~\ref{theorem:hensel-per-factor}, $d_i\hat{d}_{i+1}$}
		\State $d \gets d - \deg{f_i}$
		\EndFor
		\State $totalWork \gets \sum_{i=0}^{r-1} \mathcal{W}[i]$
		\State $ratio \gets 0$; $targ \gets 1 \,/\, t$
		\For {$i \gets 0$ to $r$}
		\State $ratio \gets ratio + \left(\mathcal{W}[i] \,/\, totalWork\right)$
		\If {$ratio \geq targ$}
		\State $\mathcal{T}[i] \gets \textsc{round}(ratio \cdot t)$;  $ratio \gets 0$
		\EndIf
		\EndFor
		\State $t \gets t_{tot} - \sum_{i=0}^{r-1} \mathcal{T}[i]$ \Comment{Give any excess threads to the earlier stages}
		\For {$i \gets 0$ to $r-1$ \textbf{while} $t > 0$}
		\State $\mathcal{T}[i] \gets \mathcal{T}[i] + 1$; $\ t \gets t - 1$
		\EndFor
%		\EndIf
		\State \Return $\mathcal{T}$
	\end{algorithmic}
\end{algorithm}

%% file: experimentation.tex
\section{Experimentation and Discussion}
\label{sec:exeripmentation}

The previous section introduced 
parallel schemes for Weierstrass preparation
and Hensel factorization based on the composition 
of the map-reduce and pipeline parallel patterns. 
Our lazy power series and parallel schemes
have been implemented in C/C++ as part of the 
Basic Polynomial Algebra Subprograms (BPAS) library \cite{bpasweb}.
These parallel algorithms are implemented using generic support
for task parallelism, thread pools, and asynchronous generators, 
also provided in the BPAS library.
The details of this parallel support are discussed
in \cite{DBLP:conf/issac/AsadiBMMX20} and \cite{DBLP:journals/jsc/AsadiBMMX20}.

Our experimentation was collected on a machine
running Ubuntu 18.04.4 with two Intel Xeon X5650 processors,
each with 6 cores (12 cores total) %(12 physical threads with hyperthreading)
at 2.67 GHz,
and a 12x4GB DDR3 memory configuration at 1.33 GHz.
All data shown is an average of 3 trials.
BPAS was compiled using GMP 6.1.2 \cite{Granlund20}.
We work over $\mathbb{Q}$ as these examples do not require algebraic numbers
to factor into linear factors. 
We thus borrow univariate integer polynomial factorization from NTL~11.4.3 \cite{shoup2020ntl}.
Where algebraic numbers are required, 
the \texttt{MultivariatePowerSeries} package of \maple \cite{DBLP:conf/maple/AsadiBKMMP20},
an extension of our work in \cite{DBLP:conf/casc/BrandtKM20}, is available.

\begin{figure}[tb]
\includegraphics[width=\textwidth]{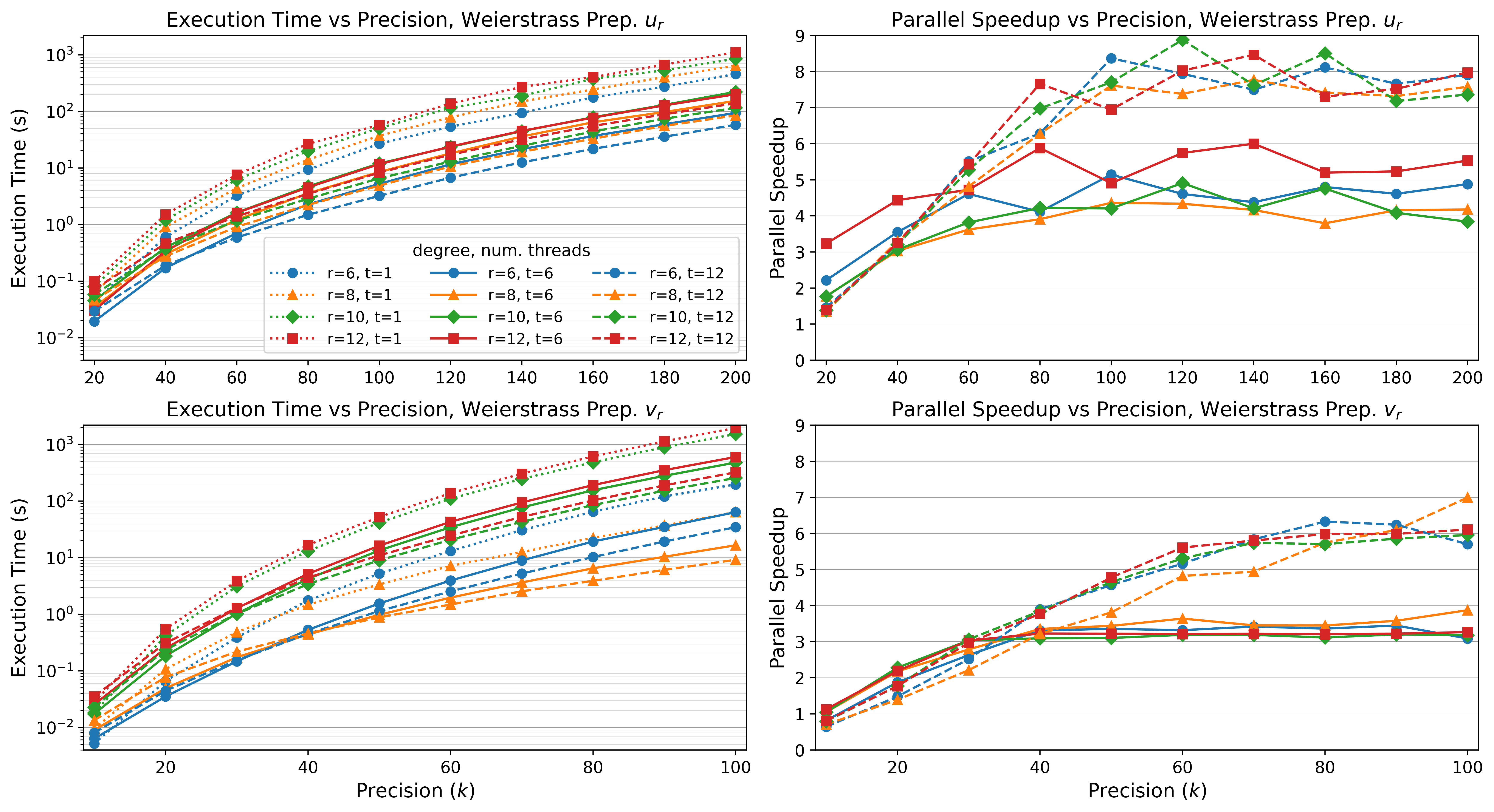}
\caption{Comparing Weierstrass preparation of $u_r$ and $v_r$ for $r \in \{6,8,10,12\}$ and number of threads $t \in \{1, 6, 12\}$. First column: execution time of $u_r$ and $v_r$; second column: parallel speedup of $u_r$ and $v_r$. Profiling of $v_6$ shows that its exceptional relative performance 
is attributed to remarkably good branch prediction.}\label{fig:weier-plots}
\end{figure}

We begin by evaluating Weierstrass preparation for two families of examples: 
\begin{enumerate}[($i$)]
	\setlength\itemsep{0.0em}
	\item $u_r = \sum_{i=2}^r (X_1^2 + X_2 + i)Y^i + (X_1^2 + X_2)Y + X_1^2 + X_1X_2 + X_2^2$
	\item $v_r = \sum_{i=\lceil r / 2 \rceil}^r (X_1^2 + X_2 + i)Y^i + \sum_{i=1}^{\lceil r /2 \rceil -1 } (X_1^2 + X_2)Y ^i + X_1^2 + X_1X_2 + X_2^2$
\end{enumerate}
Applying Weierstrass preparation to $u_r$ results in $p$ with degree 2.
Applying Weierstrass preparation to $v_r$ results in $p$ with degree $\lceil r/2 \rceil$.
Fig.~\ref{fig:weier-plots} summarizes the resulting execution times
and parallel speedups. Generally, speedup increases
with increasing degree in $Y$ and increasing precision computed.

Recall that parallelism arises in two ways:
computing summations of products of homogeneous parts
(the \parfor\ loops in Algorithms~\ref{alg:update-deg-parallel} and \ref{alg:lemma-para}),
and the \parfor\ loop over updating $c_{m-i}$ in Algorithm~\ref{alg:weierstrass-para-update2}.
The former has an inherent limitation: computing a multivariate product 
with one operand of low degree and one of high degree
is much easier than computing one where both operands
are of moderate degree. 
Evenly partitioning the iterations of the loop
does not result in even work per thread.
This is evident in comparing 
the parallel speedup between $u_r$ and $v_r$;
the former, with higher degree in $\alpha$, relies less on parallelism
coming from those products.
Better partitioning is needed and is left to future work.

We evaluate our parallel Hensel factorization
using three families of problems:
\begin{enumerate}[($i$)]
		\setlength\itemsep{0.08em}
	\item $x_r = \prod_{i=1}^r (Y-i) +  X_1(Y^3 + Y)$ 
%	\item $x_r = X_1(Y^3 + Y) + \prod_{i=1}^m (Y-i)^2$
	\item $y_r = \prod_{i=1}^r (Y-i)^i +  X_1(Y^3 + Y) $
	\item $z_r = \prod_{i=1}^r (Y + X_1 + X_2 - i) + X_1X_2(Y^3 + Y)$
\end{enumerate}
These families represent three distinct computational configurations:
$(i)$ factors of equal degree, $(ii)$ factors of distinct degrees, 
and $(iii)$ multivariate factors. 
The comparison between $x_r$ and $y_r$ is of interest in
view of Theorem~\ref{theorem:hensel-per-factor}.

\begin{figure}[tb]
	\includegraphics[width=\textwidth]{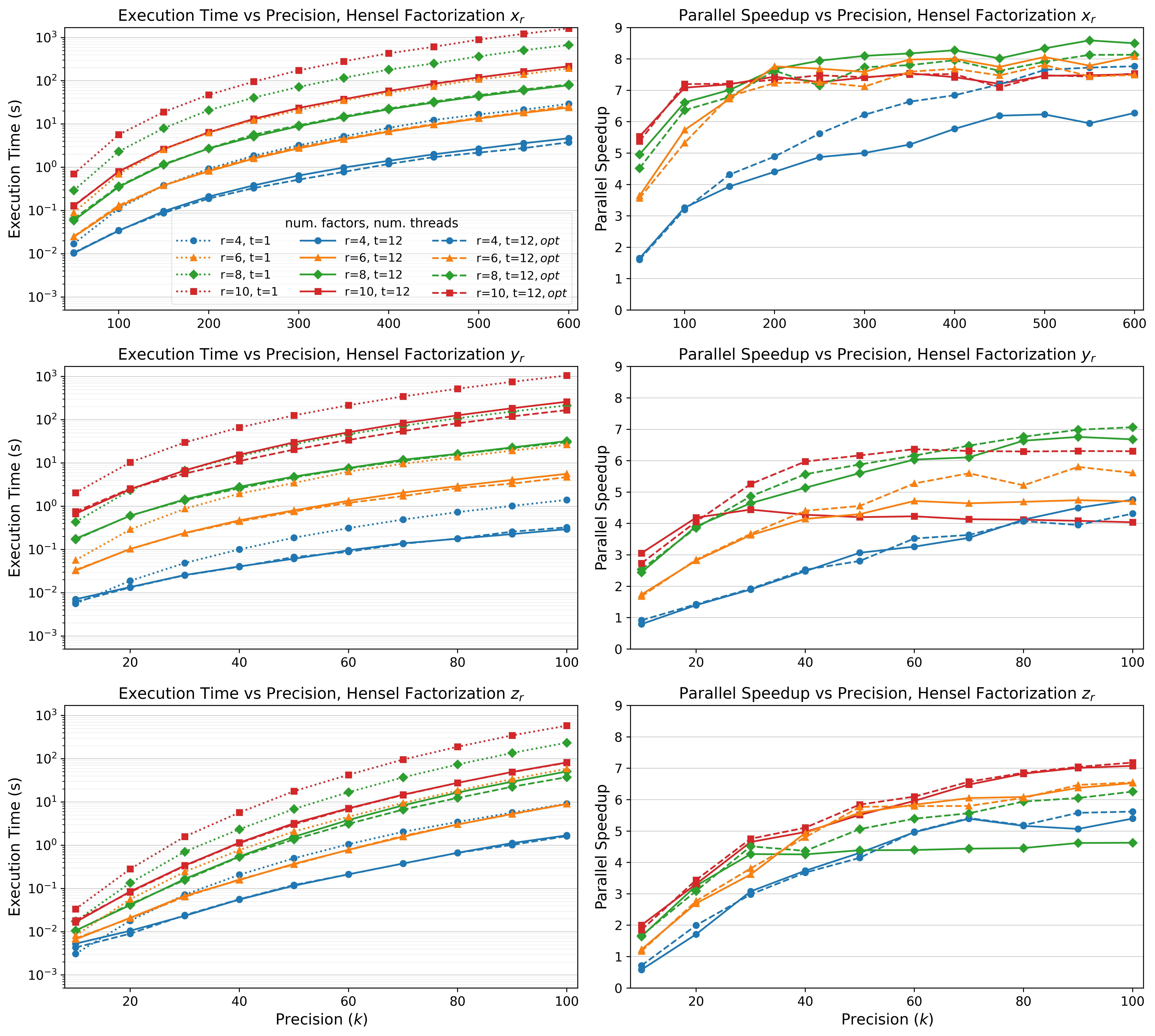}
	\caption{Comparing parallel Hensel factorization for $x_r$, $y_r$, and $z_r$ for $r \in \{4,6,8,10\}$.
		First column: execution time; second column: parallel speedup. For number of threads $t=12$ resource 
		distribution is determined by Algorithm~\ref{alg:hensel-distrib}; for $t=12,opt$ 
		serial execution time replaces complexity measures as work estimates in Algorithm~\ref{alg:hensel-distrib}, Lines~4--6.}\label{fig:hensel-plots}
\end{figure}

Despite the inherent challenges of irregular parallelism
arising from stages with unequal work, 
the composition of parallel patterns allows 
for load-balancing between stages and the overall pipeline
to achieve relatively good parallel speed-up.
Fig.~\ref{fig:hensel-plots} summarizes these results
while Table~\ref{table:hensel-time-stages} presents
the execution time per factor (or stage, in parallel).
Generally speaking, potential parallelism increases with
increasing degree and increasing precision.

The distribution of a discrete number of threads
to a discrete number of pipeline stages is a challenge;
a perfect distribution requires a fractional number
of threads~per stage. 
Nonetheless, in addition to the distribution 
technique presented in Algorithm~\ref{alg:hensel-distrib}, 
we can examine hand-chosen assignments of threads to stages. 
One can first determine the time required to update each factor
in serial, say for some small $k$, and then use that
time as the work estimates in Algorithm~\ref{alg:hensel-distrib},
rather than using the complexity estimates.
This latter technique is depicted in Fig.~\ref{fig:hensel-plots} 
as $opt$ and in Table~\ref{table:hensel-time-stages} as Time-est. threads.
This is still not perfect, 
again because of the discrete nature of threads, and
the imperfect parallelization of 
computing summations of products of homogeneous parts.
%Further research is needed to better estimate the required work
%of each pipeline stage in accounting for coefficient sizes 
%and the multivariate case.

\begin{table}[tb]
\caption{Times for updating each factor within the Hensel pipeline, 
	where $f_i$ is the factor with $i$ as the root of $\bar{f_i}$,
	for various numbers of threads per stage.
	Complexity-estimated threads use complexity estimates to estimate work 
	within Algorithm~\ref{alg:hensel-distrib};
	time-estimated threads use the serial execution time
	to estimate work and distribute threads.
%	Wait time indicates time the stage spent waiting on the previous.
	%Where a factor is assigned 0 threads,
	%it is updated along with the next factor with non-zero threads.
}\label{table:hensel-time-stages}
\centering
\scriptsize
\begin{tabular}{p{2em}lcrr|rrr|rrr}
\toprule
 &  & \multirow{2}{*}{factor$\ \ $} & \multicolumn{1}{l}{serial} & \multicolumn{1}{l}{shift} & Complexity-& \multicolumn{1}{l}{parallel}\textsl{}  & \multicolumn{1}{l}{wait}  & Time-est.& \multicolumn{1}{l}{parallel}\textsl{}  & \multicolumn{1}{l}{wait}\\
 &  &       & \multicolumn{1}{l}{time (s)}  &  \multicolumn{1}{l}{time (s)}  & Est. threads  & time (s)  & \multicolumn{1}{l}{time (s)} &  threads  & time (s)  & \multicolumn{1}{l}{time (s)}           \\
\midrule
$x_4$ & $k$ = 600 & $f_1$ & 18.1989 & 0.0012 & 6 & 4.5380 & 0.0000 & 7 & 3.5941 & 0.0000\\
       &     & $f_2$ & 6.6681 & 0.0666 & 4 & 4.5566 & 0.8530 & 3 & 3.6105 & 0.6163\\
       &     & $f_3$ & 3.4335 & 0.0274 & 1 & 4.5748 & 1.0855 & 0 & - & -\\
       &     & $f_4$ & 0.0009 & 0.0009 & 1 & 4.5750 & 4.5707 & 2 & 3.6257 & 1.4170\\
%       &     & $f_5$ & - & - & - & - & - & - & - & -\\
       \cline{4-11} 
       \multicolumn{2}{c}{\textit{totals}$\quad$} &  & 28.3014 & 0.0961 & 12 & 4.5750 & 6.5092 & 12 & 3.6257 & 2.0333\\
\midrule
$y_4$ & $k$ = 100 & $f_1$ & 0.4216 & 0.0003 & 3 & 0.1846 & 0.0000 & 4 & 0.1819 & 0.0000 \\
&      & $f_2$ & 0.5122 & 0.0427 & 5 & 0.2759 & 0.0003 & 4 & 0.3080 & 0.0001 \\
&      & $f_3$ & 0.4586 & 0.0315 & 3 & 0.2842 & 0.0183 & 0 & - & - \\
&      & $f_4$ & 0.0049 & 0.0048 & 1 & 0.2844 & 0.2780 & 4 & 0.3144 & 0.0154 \\
%&     & - & - & - & - & - & - & - & - & -\\
\cline{4-11} 
\multicolumn{2}{c}{\textit{totals}$\quad$}  &  & 1.3973 & 0.0793 & 12 & 0.2844 & 0.2963 & 12 & 0.3144 & 0.0155 \\
\midrule
$z_4$ & $k$ = 100  & $f_1$ & 5.2455 & 0.0018 & 6 & 1.5263 & 0.0000 & 7 & 1.3376  & 0.0000\\
&      & $f_2$ & 2.5414 & 0.0300 & 4& 1.5865 & 0.2061  & 3 & 1.4854 & 0.0005 \\
&      & $f_3$ & 1.2525 & 0.0151 & 1& 1.6504 & 0.1893  & 0 & - &  -\\
&      & $f_4$ & 0.0018 & 0.0018 & 1& 1.6506 & 1.6473 & 2 & 1.5208 &  0.7155\\
%&     & - & - & - & - & - & - & - & - & -\\
\cline{4-11} 
\multicolumn{2}{c}{\textit{totals}$\quad$} &  & 9.0412 & 0.0487 & 12 & 1.6506 & 2.0427 & 12 & 1.5208 & 0.7160 \\
\bottomrule
\end{tabular}
\end{table}

In future, we must consider several important factors to improve
performance. Relaxed algorithms should give better complexity and performance.
For parallelism, better partitioning schemes for the map-reduce
pattern within Weierstrass preparation 
should be considered. Finally, for the Hensel pipeline, 
more analysis is needed to optimize the scheduling
and resource distribution, particularly considering
coefficient sizes and the multivariate case.